\documentclass[11pt,a4paper]{article}
\usepackage[utf8]{inputenc}
\usepackage[T1]{fontenc}
\usepackage[english]{babel}
\usepackage{amsmath,amssymb,amsfonts,amsthm}
\usepackage{graphicx}\graphicspath{{fig/}}
\usepackage{geometry}
\usepackage{booktabs}
\usepackage{algorithm}
\usepackage{algpseudocode}
\usepackage{tikz}
\usetikzlibrary{positioning}
\usepackage{url}
\usepackage[hidelinks]{hyperref}
\DeclareUnicodeCharacter{03C9}{\ensuremath{\omega}}
\DeclareUnicodeCharacter{03BB}{\ensuremath{\lambda}}
\DeclareUnicodeCharacter{0394}{\ensuremath{\Delta}}
\DeclareUnicodeCharacter{221E}{\ensuremath{\infty}}

\newcommand{\dgm}{\mathrm{dgm}}
\newcommand{\DGM}{\mathcal{D}}
\newcommand{\R}{\mathbb{R}}
\newcommand{\N}{\mathbb{N}}
\newcommand{\dW}{d_{\mathcal W}}
\newcommand{\bb}{b}
\newcommand{\dd}{d}
\DeclareMathOperator*{\argmin}{arg\,min}
\DeclareMathOperator*{\argmax}{arg\,max}

\newtheorem{proposition}{Proposition}

\theoremstyle{remark}
\newtheorem{remark}{Remark}

\title{\textbf{Active flow control based on topological data analysis and a DMDc reduced-order model for drag reduction}}
\author{C. Mbangue Lobe Yiphta and D. Razafindralandy}
\date{Laboratoire des Sciences de l'Ingénieur pour l'Environnement (LaSIE)\\
La Rochelle Université, France}

\begin{document}

\maketitle

\begin{abstract}
Controlling and reducing undesirable vortical structures in unsteady flows is a major issue in aerodynamics. This paper presents an active control methodology that combines topological data analysis (TDA) through persistence diagrams, a reduced-order model based on dynamic mode decomposition with control (DMDc) and receding-horizon model predictive control (MPC). Vortices are detected and characterised by the persistence diagram of the absolute value of the vorticity, which makes it possible to define a target flow in which only the useful vortices are retained. This target flow is computed offline in the reduced space by gradient descent, the derivative of a persistence diagram being defined through a parametrisation of the space of diagrams. The control loop then reduces to a small-dimensional convex quadratic program, whose construction we detail and whose cost we compare with that of an approach that places the diagrams directly in the cost function. The approach is applied numerically to the flow around an oscillating aircraft wing for seven vibration frequencies, and yields a mean reduction of the drag coefficient of about $82\,\%$ while preserving the main vortices.
\medskip

\end{abstract}

\section{Introduction}\label{sec:intro}

Drag reduction and flow optimisation around streamlined bodies (aircraft wings, road vehicles, \ldots) rely largely on the control of vortical structures. Depending on their nature and position, vortices have opposite effects: the large vortex near the leading edge of a wing increases lift, whereas other structures, notably in the wake, are strongly dissipative and cause substantial energy losses. The aim is therefore not to suppress all vortices, but to be able to detect and characterise them, and then to act selectively on the undesirable ones. There are also situations where, on the contrary, one seeks to create vortices, for instance to delay stall, enhance fluid mixing or accelerate transfers; these cases are not considered here.

Classical active control strategies rely on global cost functions (kinetic energy, enstrophy, drag) evaluated in the full physical space, and on models obtained by discretising the Navier--Stokes equations. The high dimension of these models and the non-convexity of the associated optimisation problems make real-time application difficult. Two ingredients partly overcome these difficulties: \emph{model reduction}, which replaces the flow by a small number of modal coefficients, and \emph{model predictive control} (MPC), which recasts control as a sequence of finite-horizon optimisation problems. The choice of the reduced model is, however, decisive: a Galerkin projection leads to nonlinear dynamics and non-convex costs, whereas a linear model identified from data, such as dynamic mode decomposition with control (DMDc), leads to a convex quadratic program.

While POD--DMDc--MPC makes flow control tractable, it does not by itself describe which vortical structures are present or should be preserved. Persistent homology, a branch of topological data analysis (TDA) grounded in algebraic topology, provides this complementary structural description by characterising qualitative features of data through Betti numbers, which count connected components, holes and cavities. Tracked across scales, these invariants are summarised in a persistence diagram, a compact topological signature whose stability makes it robust to small perturbations in noisy or incomplete simulation and experimental data. In fluid mechanics, persistence diagrams built from vorticity or enstrophy have been used to characterise turbulent patterns and compare flow fields \cite{Vivodtzev2024TDA}; TDA has also characterised particle trajectories in concentrated fibre suspensions \cite{Mezher2021TDA}, monitored coherent vortices in dynamic stall \cite{Martinez2026}, and identified periodic or recurrent structures in Kolmogorov flow and Rayleigh--Bénard convection \cite{Kramar2015Kolmogorov}. Other applications include estimating the permeability of fracture networks directly from images \cite{Suzuki2021Flow}, describing transitions between periodic and chaotic regimes from evolving streamline patterns \cite{Sakajo2026TFDA}, and combining persistent homology with reduced-order dynamics and data assimilation to forecast convective systems \cite{Canot2026TDT}. These applications illustrate how TDA complements energy-based measures with structural information. More broadly, it could help assess which topological features a POD reduction preserves and provide a posteriori indicators of its reliability, but its use to guide flow control remains comparatively unexplored. Here, POD supplies a low-dimensional representation, persistence diagrams characterise its vortical structures and define the target, and the linear DMDc model provides predictions that MPC uses to steer the flow toward that target.

In this paper, we first propose a topological definition of the \emph{target flow}, based on the persistence diagram of the vorticity, and its offline computation in the reduced space. To solve this computation by gradient descent, we compute the derivative of a persistence diagram with respect to the modal coefficients. We then build an MPC controller with augmented state acting on the angle of attack, which reduces to a convex quadratic program, and we justify the offline computation of the target state by a computational-cost analysis. Finally, we validate the approach on the flow around an oscillating wing, for seven frequencies, with a mean drag reduction of about $82\,\%$.

The remainder of the paper is organised as follows. Section~\ref{sec:prelim} recalls the Navier--Stokes equations, the reduced model based on proper orthogonal decomposition (POD) and DMDc, and section~\ref{sec:persistance} recalls persistent homology. Sections~\ref{sec:cible} to~\ref{sec:precalc_a_star} present the method, section~\ref{sec:resultats} the numerical results, and section~\ref{sec:discussion} discusses limitations and perspectives.

\section{Framework: flow and model reduction}\label{sec:prelim}

\subsection{Governing equations and the control problem}
The dynamics of an incompressible viscous fluid is governed by the Navier--Stokes equations together with the incompressibility constraint:
\begin{equation}
	\frac{\partial u}{\partial t} + (u \cdot \nabla)u = -\nabla p + \nu \Delta u + f, \qquad \nabla \cdot u = 0,
	\label{eq:ns}
\end{equation}
where $u$ is the velocity field, $p$ the pressure, $\nu$ the kinematic viscosity and $f$ the external forces. In active control, one acts on a control parameter, here the angle of attack $\alpha$ of a wing, in order to modify the flow dynamics and reduce the impact of parasitic vortices. For simplicity, we restrict ourselves to the two-dimensional case.

\subsection{POD reduced model}
From snapshots of the velocity field, a proper orthogonal decomposition (POD) \cite{lumley1970stochastic} provides a basis in which the reduced velocity field reads
\[
u_r(x,t)=\sum_{i=1}^{r} a_i(t)\,\phi_i(x),
\]
with $r\ll N$, $N$ being the number of spatial discretisation points. The flow is thus represented by the vector of coefficients $a(t)\in\R^r$. The vorticity of the reduced field is denoted $\omega(a)$ to emphasise its dependence on $a$.

\subsection{Limits of Galerkin projection and the benefit of DMDc}\label{sec:dmdc}
The classical reduced model (POD--Galerkin) is obtained by projecting \eqref{eq:ns} onto the POD basis; it leads to nonlinear dynamics for $a$. In optimal control, such dynamics generally associate with the cost function a \emph{non-convex} problem: it may have local minima, and its solution requires iterative methods (adjoint computation, gradient, dynamic programming) with no guarantee of convergence to the global minimum.

Dynamic mode decomposition (DMD) \cite{schmid2010dynamic} and its version with control (DMDc) \cite{proctor2016dynamic} identify a \emph{linear} model from snapshots. For DMDc, the dynamics of the reduced state reads
\begin{equation}
	a(t+\delta t) = A\, a(t) + B\, \boldsymbol{\alpha}(t),
	\label{eq:dmdc}
\end{equation}
where $A \in \R^{r \times r}$ and $B = [B_\alpha \mid B_{\dot{\alpha}}] \in \R^{r \times 2}$ are the model matrices, computed from the snapshots, and $\boldsymbol{\alpha}=(\alpha,\dot{\alpha})^\top$ gathers the angle of attack and its time derivative. Besides its speed, the linearity of \eqref{eq:dmdc} is an essential mathematical property: it turns the evolution constraints over the prediction horizon into affine constraints on the control variables, and a quadratic cost into a convex problem. In the absence of constraints on the control variables, the optimal control law is even obtained by solving a Riccati equation, which has a unique analytical solution.

\section{Persistent homology and persistence diagrams}\label{sec:persistance}

This section introduces essential tools from persistent homology for analysing functions. In our applications, the function is chosen to characterise vortices. It may, for example, be the $Q_2$ criterion, the $\lambda_2$ function or the $\Delta$ function \cite{jeong1995identification,hunt1988eddies,zhou1999mechanisms,zhang2018review,qu2024vortex}. To obtain the topological signature of a function, we determine the topological features of its sublevel sets.

\subsection{Persistence barcode and diagram from sublevel sets}

Let $g:\Omega\to \mathbb R$ be a function. Consider its sublevel sets
\[
X^s=\{x\in\Omega \mid g(x)\le s\}\qquad s\in\R .
\]
Sweeping $s$ from $-\infty$ to $+\infty$ amounts to sweeping the level of $|\omega|$ from its highest to its lowest values. The set $X^s$ grows with $s$, forming a \emph{filtration} of $\Omega$.

\begin{figure}[ht]\centering
	\includegraphics[width=.4\textwidth]{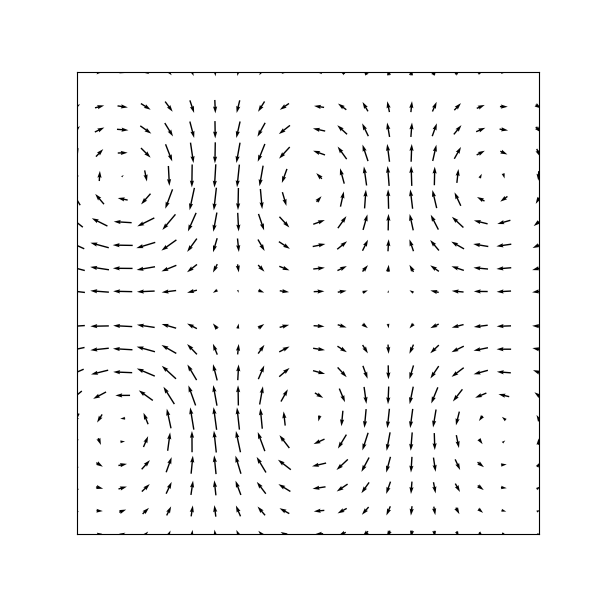}
	\includegraphics[width=.4\textwidth]{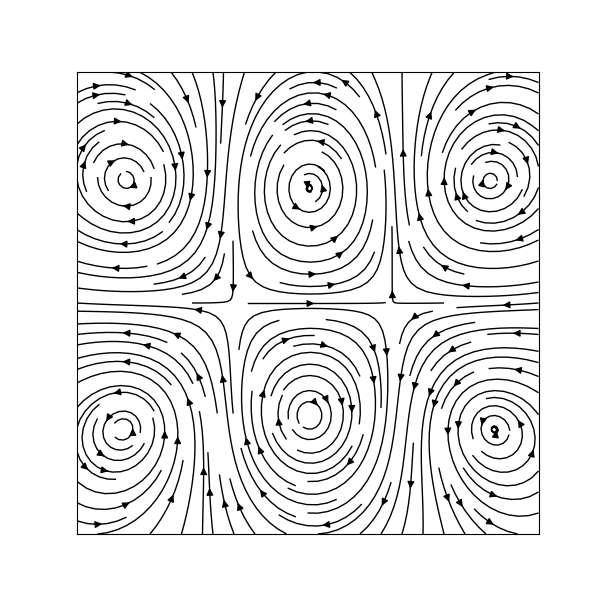}
	\caption{Velocity field and streamlines}
	\label{fig:champ_vitesse}
\end{figure}

\begin{figure}[ht]
	\centering
	\hfill
	\includegraphics[width=.4\textwidth]{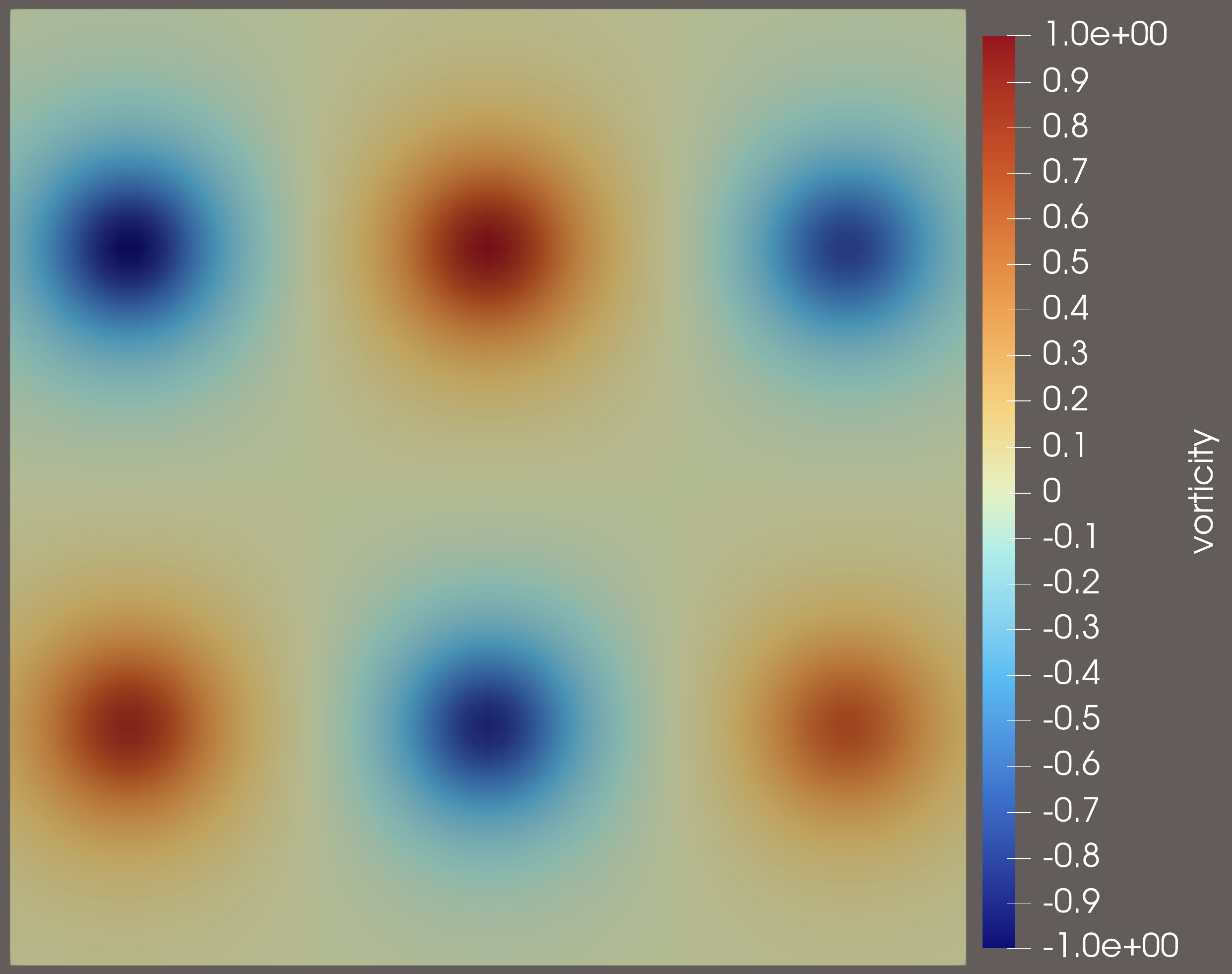}
	\hfill
	\includegraphics[width=.4\textwidth]{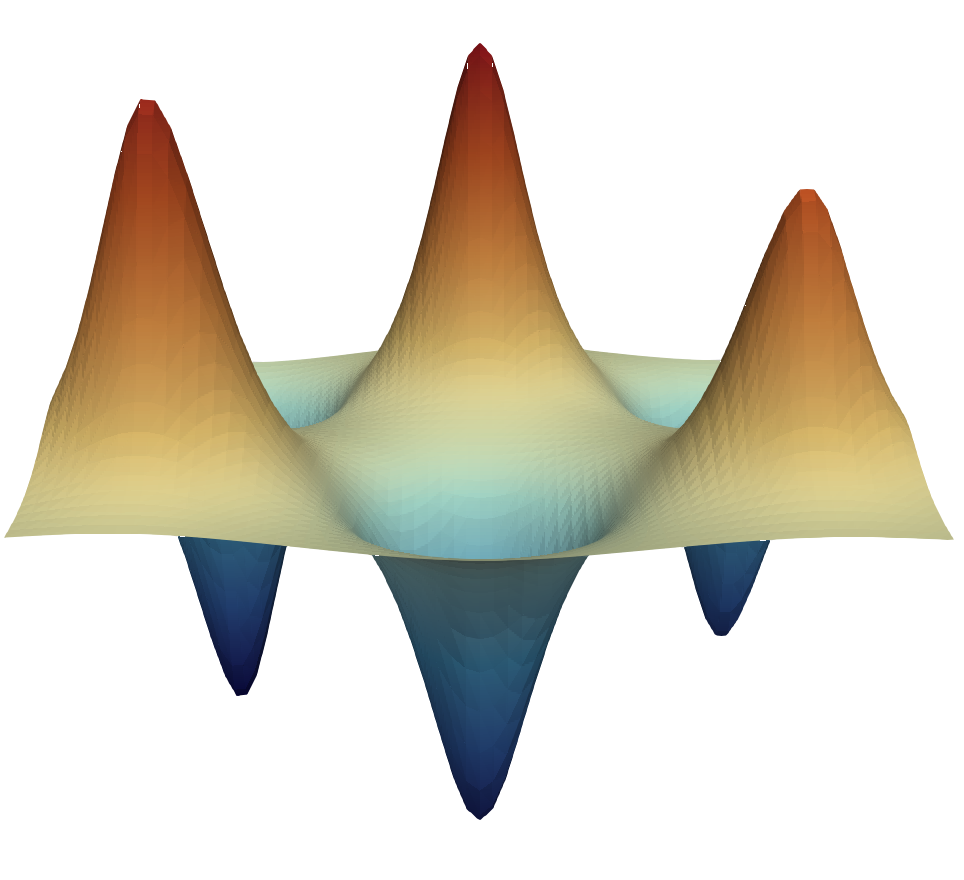}
	\hfill
	\caption{Vorticity field shown as a colormap (left) and a surface (right)}
	\label{fig:champ_vorticite}
\end{figure}

Consider for example the two-dimensional flow shown in figure \ref{fig:champ_vitesse}, where the velocity field is displayed with several streamlines. This theoretical flow is generated by superposing six Scully vortices with different rotation directions and intensities ranging from $0.7$ to $1$. We take the vorticity field $\omega$, shown in figure \ref{fig:champ_vorticite}, as the filtration function $g$.

\begin{figure}[ht]\scriptsize
	\centering
	\begin{tabular}{cccc}
		\includegraphics[width=.2\textwidth]{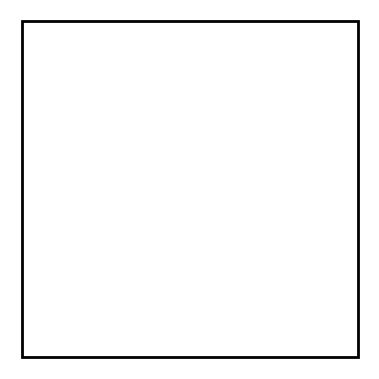}&
		\includegraphics[width=.2\textwidth]{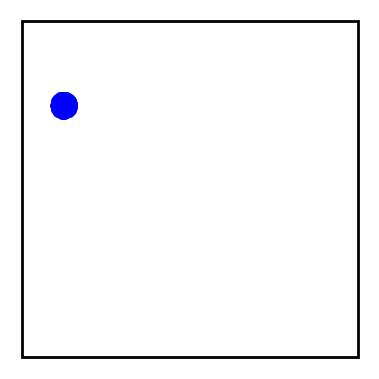}&
		\includegraphics[width=.2\textwidth]{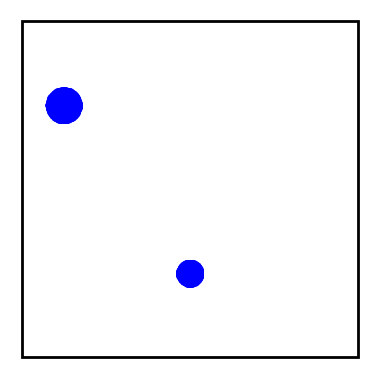}&
		\includegraphics[width=.2\textwidth]{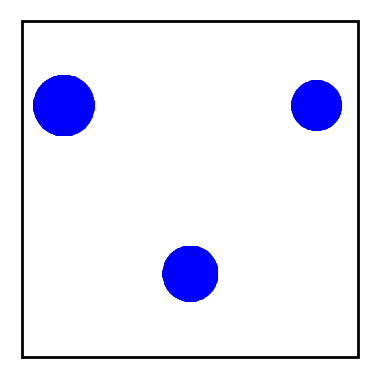}\\
		$s=-1.1$ & $s=-0.9$ & $s=-0.8$ & $s=-0.5$\\
		\includegraphics[width=.2\textwidth]{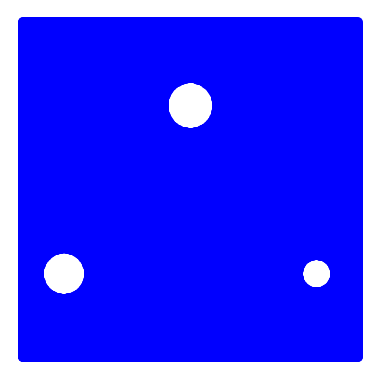}&
		\includegraphics[width=.2\textwidth]{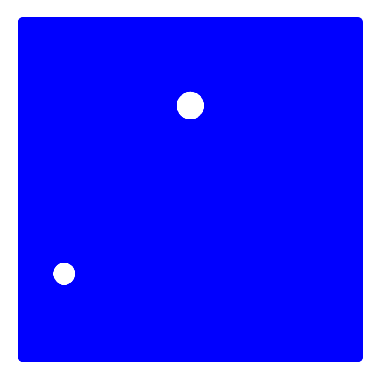}&
		\includegraphics[width=.2\textwidth]{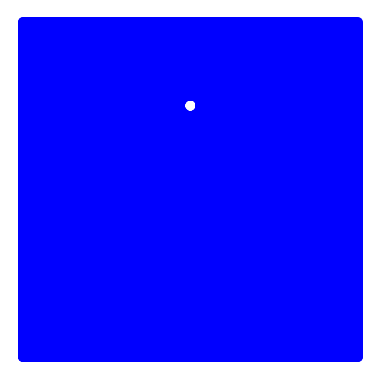}&
		\includegraphics[width=.2\textwidth]{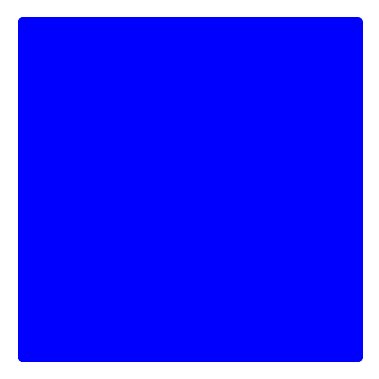}\\
		$s=0.5$ & $s=0.7$ & $s=0.9$ & $s=1.1$\\
	\end{tabular}
	\caption{The set $X^s$ for several threshold values $s$.}
	\label{fig:filtration_sous_niveaux}
\end{figure}

\begin{figure}[htp]
	\centering
	\includegraphics[width=.4\textwidth]{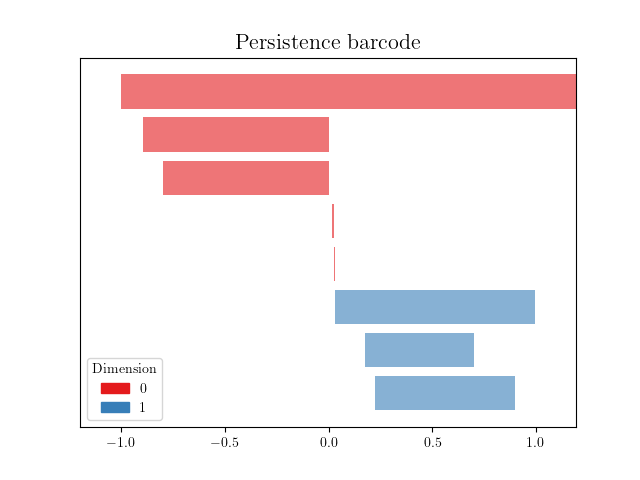}
	\includegraphics[width=.4\textwidth]{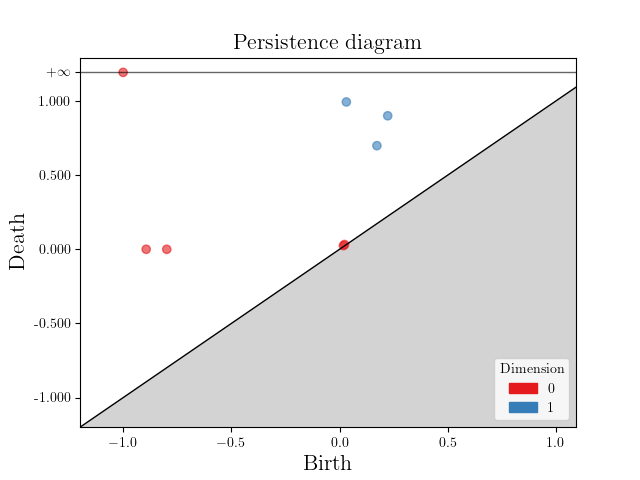}
	\caption{Barcode and persistence diagram $\dgm(\omega)$ associated with the vorticity. Dimension-$0$ homology is shown in red and dimension-$1$ homology in blue.
	}
	\label{fig:barcode_diagramme_persistence}
\end{figure}

The sets $X^s$ for several values of $s$ are shown in figure \ref{fig:filtration_sous_niveaux}. Since the vorticity ranges from $-1$ to $1$, the set $X^s$ is empty for $s < -1$. As $s$ increases, connected components appear: one is born at approximately $b=-1$ while two others are born at approximately $b=-0.89$ and $b=-0.79$. These two last components merge into the main component at the threshold $s\simeq 0$. Consequently, we have the persistence the intervals $(-0.89,0)$ and $(-0.78,0)$ for the two last components and $(-1,+∞)$ for the main one. They constitute the long red bars in the barcode of figure \ref{fig:barcode_diagramme_persistence}. Some very short red bars can also be seen on this figure, corresponding to numerical discretisation noise. These interval represent the persistence of the Betti numbers of the $X^s$ of dimension 0.

We now consider homology of dimension 1. For $s<0$, the sublevel set $X^s$ contains no holes. As $s$ increases from $0$ to $0.5$, three holes appear in succession; as the threshold keeps increasing, they are filled one by one. Their birth--death intervals are $(0.31,0.99)$, $(0.17,0.69)$ and $(0.22,0.89)$, represented by the blue bars in figure \ref{fig:barcode_diagramme_persistence}.



The intervals $(b,d)$ form the persistence barcode of $g$. Since points are easier to handle than intervals, we use the persistence diagram, which is the multiset of points
$(b,d)\in\R\times\,]-\infty,+\infty]$.

For the theoretical flow in figure \ref{fig:champ_vitesse}, the diagram $\dgm(\omega)$ counts the vortices and distinguishes them by their persistence and rotation direction. To analyse vortices without taking their rotation direction into account, one can consider the diagram $\dgm(|\omega|)$ of the absolute value of the vorticity. In this case, it is the 1-dimensional homology which is relevant, since each vortex is then represented by a hole in the sublevel sets of $|\omega|$.

\begin{remark}
The present construction is an instance of a general theory. The filtration induces a \emph{persistence module}, which decomposes uniquely into \emph{interval modules}; the endpoints of these intervals are exactly the pairs $(b,d)$ defined above. This guarantees that the diagram is well defined and independent of any choice made to track the components \cite{edelsbrunner2010computational,chazal2016structure}.
\end{remark}

\subsection{Comparing persistence diagrams}

To compare diagrams, that may contain different numbers of points, partial matchings are used: a \emph{partial matching} between two diagrams $D$ and $D'$ is a subset $C\subset D\times D'$ whose two projections $\pi_1$ and $\pi_2$ are injective. Unmatched points $p$ are paired with their orthogonal projection $\pi_{Δ}(p)$ to the diagonal $\Delta$ of $\mathbb R^2$. The $q$-cost of $C$, for $q\in\N^*$, and the $\infty$-cost are respectively
\[
c_q(C)=\Big(\sum_{(p,p')\in C}\|p-p'\|_\infty^q+\sum_{p\in D\setminus\pi_1(C)}\|p-\pi_\Delta(p)\|_\infty^q+\sum_{p'\in D'\setminus\pi_2(C)}\|p'-\pi_\Delta(p')\|_\infty^q\Big)^{1/q},
\]
\[
c_{∞}(C)=\max_{(p,p')\in C}\|p-p'\|_\infty+\max_{p\in D\setminus\pi_1(C)}\|p-\pi_\Delta(p)\|_\infty+\max_{p'\in D'\setminus\pi_2(C)}\|p'-\pi_\Delta(p')\|_\infty.
\]
The \emph{Wasserstein distance of order $q$} and the  \emph{bottleneck distance} between $D$ and $D'$ are respectively
\[d_q(D,D')=\inf_C c_q(C),\qquad d_\infty(D,D')=\inf_C c_\infty(C),\] the infima being taken over all partial matchings.
In particular, $d_q(D,\dgm(0))=\big(\sum_{p\in D}\|p-\pi_\Delta(p)\|_\infty^q\big)^{1/q}$ is the cumulated persistence of $D$, where $\dgm(0)=\emptyset$ is the diagram of the null field. Generally, infinite points are removed from the diagrams before computing distances to keep a finite cost.

An essential property of persistence diagrams is their stability: a small perturbation of a function only slightly changes its diagram. More precisely, if $g,h:\Omega\to\R$ are two continuous (tame) functions \cite{cohensteiner07,chazal2014persistence}, then
\[d_\infty\big(\dgm(g),\dgm(h)\big)\le\|g-h\|_\infty.\]

\section{Target flow}\label{sec:cible}

To build a control algorithm, one must first define the flow one wishes to obtain: a flow in which the undesirable vortices are reduced or eliminated. In the context of a flow around a wing, assume that we wish to keep the main vortex near the leading edge, which is beneficial for lift, and to reduce the other vortices in the wake, which are dissipative. The target flow is therefore a flow containing only the main vortex. In other contexts, one may wish to keep several vortices, for instance the main vortex and a secondary vortex.

For simplicity, vortices are identified from the vorticity of the field, as in the previous section. For a more robust analysis of vortices in three-dimensional flows, criteria such as $Q_2$, $\lambda_2$, $Q$ or $\Delta$ are generally preferable \cite{jeong1995identification,hunt1988eddies,zhou1999mechanisms,zhang2018review,qu2024vortex}.

\subsection{Target persistence diagram}

With the POD framework with a fixed basis, finding the target flow amounts to finding the coefficients $a=(a_1,\dots,a_r)$ of the reduced flow $u_r$ that correspond to the target. Let $\omega(a)$ denote the vorticity field associated with $u_r$, and let $\dgm(|\omega(a)|)$ denote the persistence diagram of the function $|\omega(a)|$. The target flow is such that its persistence diagram contains only the most persistent point, corresponding to the vortex to be retained. We denote by $\dgm'(|\omega(a)|)$ the diagram obtained by removing this point, assuming it is unique:
\begin{equation}
	\dgm'(|\omega(a)|) = \dgm(|\omega(a)|) \setminus \Big\{\argmax_{(\bb,\dd)\in\dgm(|\omega(a)|)}(\dd-\bb)\Big\}.
	\label{eq:dgm_prime}
\end{equation}
The target flow is a flow containing only the vortex to keep: for it, $\dgm'(|\omega(a)|)=\emptyset=\dgm(0)$. The flow must therefore be controlled so as to minimise the distance $d\big(\dgm'(|\omega(a)|),\dgm(0)\big)$, where $d$ is for instance the Wasserstein distance $d_2$.

Definition \eqref{eq:dgm_prime} can be adapted to the application: one may remove the $k$ most persistent points, or the points whose persistence exceeds a threshold. In the simulations of section~\ref{sec:resultats}, we remove the two most persistent points: the control then aims at reducing all the vortices other than the main vortex and the secondary vortex.
Note that a persistence diagram contains no positional information: it cannot be used to select a vortex according to its location (leading-edge vortex, wing-tip vortex, \ldots). To do so, it must be combined with other information, for instance the mergegram \cite{elkin2020mergegram}.

\subsection{Offline computation of the target state}
Introducing $d(\dgm'(|\omega(a)|),\dgm(0))$ directly in the controller cost function would require reconstructing the field $\omega(a)\in\R^N$ at each evaluation, and would destroy convexity (section~\ref{sec:precalc_a_star}). We therefore compute, once and offline, the reduced coefficient $a^*$ that minimises this distance:
\begin{equation}
	a^* = \argmin_{a\in\R^r} d_{2}\big(\dgm'(|\omega(a)|),\dgm(0)\big).
	\label{eq:a_star}
\end{equation}
A bottleneck distance is also conceivable, as are vector representations of the diagrams (persistence images, \ldots). Several methods exist to solve \eqref{eq:a_star} \cite{boyd2004convex,nocedal2006numerical,rios2013derivative,katoch2021review}; we have chosen gradient descent. This requires computing the gradient of the distance and, by composition, the derivative of a persistence diagram with respect to $a$, which is the subject of the next subsection.

\subsection{Derivative of a persistence diagram}\label{sec:derivee}
The set $\DGM$ of persistence diagrams is a metric space, but is neither a vector space nor even a manifold: the derivative therefore cannot be defined in the classical sense. However, $\DGM$ can be endowed with a \emph{diffeology}, which generalises the notion of manifold \cite{souriau2006groupes,souriau2005groupes,iglesias2017introduction,leygonie2022framework}. We retain only the essentials needed to compute the gradient.

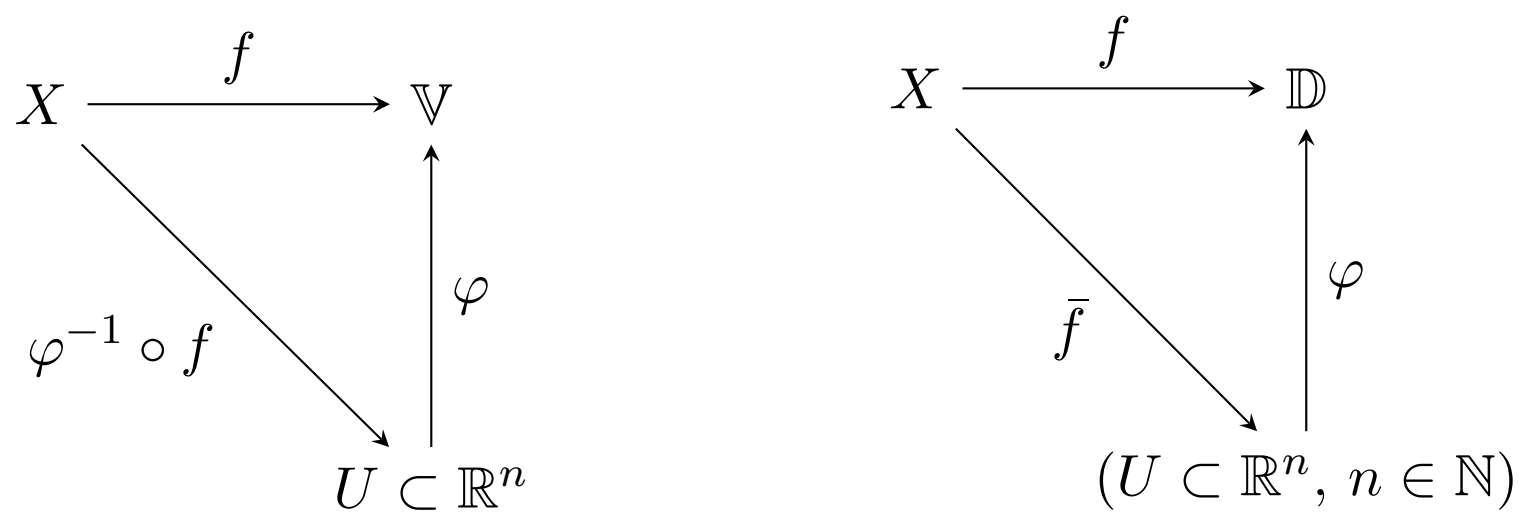
\begin{figure}\centering
	\begin{tikzpicture}[>=stealth, node distance=2cm]
		\node (a) {$X$};
		\node (b) [right=of a] {$\mathbb V$};
		\node (c) [below=of b] {$U\subset\R^n$};
		\draw[->] (a) --node[above]{$f$} (b);
		\draw[->] (a) --node[below left]{$\varphi^{-1}\circ f$} (c);
		\draw[->] (c) --node[right] {$\varphi$} (b);
	\end{tikzpicture}\hspace{2cm}
	\begin{tikzpicture}[>=stealth, node distance=2cm]
		\node (a) {$X$};
		\node (b) [right=of a] {$\mathbb D$};
		\node (c) [below=of b] {$(U\subset\R^n,\,n\in\N)$};
		\draw[->] (a) --node[above]{$f$} (b);
		\draw[->] (a) --node[below left]{$\bar f$} (c);
		\draw[->] (c) --node[right] {$\varphi$} (b);
	\end{tikzpicture}
	\caption{Left, the case of a manifold $\mathbb V$: $(U,\varphi)$ is a local chart and $f$ is differentiable if it has a differentiable local representation $\varphi^{-1}\circ f$. Right, the case of a diffeological space $\mathbb D$: $(U,\varphi)$ is a parametrisation and $f$ is differentiable if it has a differentiable lift $\bar f$ ($\bar f$ is a lift of $f$ if there exists a parametrisation $(U,\varphi)$ such that $f=\varphi\circ\bar f$).}
	\label{fig:diffeologie}
\end{figure}

\paragraph{Diffeology.} A diffeology on a set $\mathbb D$ is a family of functions $\varphi:U\subset\R^n\to\mathbb D$ satisfying certain compatibility conditions. The pairs $(U,\varphi)$ define parametrisations of $\mathbb D$ and generalise the local charts of a manifold (figure~\ref{fig:diffeologie}). Unlike a local chart, the dimension $n$ of the open set $U$ may vary from one parametrisation to another, which makes $\mathbb D$ a stratified space. The differentiability and derivative of a function $f$ from a manifold $X$ to $\mathbb D$ are then reduced to those of a lift $\bar f$ (the derivative depends on the parametrisation).

\paragraph{Application to diagrams.} The space $\DGM$ is naturally stratified:
\[
\DGM = \bigcup_{m,n\ge 0} \DGM_{2m+n},
\]
where $\DGM_{2m+n}$ is the subset of diagrams containing exactly $m$ finite points and $n$ points with infinite ordinate:
\[
\DGM_{2m+n} \subset \big\{ (\bb_1,\dd_1),\ldots,(\bb_m,\dd_m),(\bb_{m+1},+\infty),\ldots,(\bb_{m+n},+\infty)\big\}\simeq\R^{2m+n}.
\]
The inclusions $\DGM_{2m+n}\subset\R^{2m+n}$ provide the parametrisations. The derivative of a diagram
\[
D(x)= \big\{(\bb_1(x),\dd_1(x)),\ldots,(\bb_m(x),\dd_m(x)),(\bb_{m+1}(x),+\infty),\ldots,(\bb_{m+n}(x),+\infty)\big\}
\]
with respect to a parameter $x$ is defined by the derivatives of the $\bb_i(x)$ and $\dd_i(x)$ with respect to $x$. This derivative is not unique: it depends on the ordering of the points. In practice, the points are ordered by increasing birth, then by increasing death; the parametrisation, and hence the derivative, are then unique except in the presence of multiple points. In our problem, $x=a$ is the vector of coefficients in the POD basis: $\bb_i$ and $\dd_i$ are values taken by $|\omega(a)|$ at critical points, and depend on $a$ through $\omega(a)$, which is linear in $a$. The derivative of the distance $d_2(\dgm'(|\omega(a)|),\dgm(0))$ with respect to $a$ is then obtained by composition, which enables gradient descent for \eqref{eq:a_star}.

\section{Model predictive control (MPC)}\label{sec:mpc}

Once the target flow, characterised by $a^*$, is defined, one must act on the control parameter to reach it. For the flow around a wing, we take the angle of attack $\alpha$ as the control parameter. We use model predictive control (MPC) \cite{garcia1989model,schwenzer2021review}: the evolution of the flow is predicted over a future horizon, and the control is chosen to minimise the gap between the predicted trajectory and the target flow.

\subsection{Prediction model and augmented state}
The prediction relies on the DMDc model \eqref{eq:dmdc}, parametrised by $\alpha$ and $\dot\alpha$. These two quantities are not independent: they are linked by the kinematic relation $\dot\alpha=d\alpha/dt$, which, discretised with a first-order explicit Euler scheme, becomes
\begin{equation}
	\alpha(t+\delta t) = \alpha(t) + \delta t\, \dot{\alpha}(t).
	\label{eq:cinematique_discrete}
\end{equation}
Relations \eqref{eq:dmdc} (flow dynamics) and \eqref{eq:cinematique_discrete} (actuator kinematics) must be satisfied simultaneously. We therefore introduce the \emph{augmented state}
\[
\tilde{x}(t) = \begin{bmatrix} a(t) \\ \alpha(t) \end{bmatrix} \in \R^{r+1},
\]
which obeys the linear system
\begin{equation}
	\tilde{x}(t+\delta t) = \tilde{A}\, \tilde{x}(t) + \tilde{B}\, \dot{\alpha}(t),
	\label{eq:augmente}
\end{equation}
with
\[
\tilde{A} = \begin{bmatrix} A & B_\alpha \\ 0_{1\times r} & 1 \end{bmatrix} \in \R^{(r+1)\times(r+1)}, \qquad
\tilde{B} = \begin{bmatrix} B_{\dot{\alpha}} \\ \delta t \end{bmatrix} \in \R^{(r+1)\times 1}.
\]
The first row of $\tilde A$ retains the DMDc dynamics with the effect of $\alpha$ on $a$, while the second row expresses the integration of $\alpha$ through its increment $\delta t\,\dot\alpha$. The matrix $\tilde B$ only contains the effect of $\dot\alpha$ on the physical state (first row) and its integrating effect on $\alpha$ (second row). The quantity $\dot\alpha(t)$ thus becomes the only control variable.

\subsection{Optimisation problem}
For a horizon of $n_t$ time steps, the prediction is given by the recurrence $\tilde{x}(t+(n+1)\delta t) = \tilde{A}\,\tilde{x}(t+n\delta t) + \tilde{B}\,\dot\alpha(t+n\delta t)$, $n=0,\ldots,n_t-1$. We look for the controls $\dot\alpha^*(t),\ldots,\dot\alpha^*(t+(n_t-1)\delta t)$ that minimise
\begin{equation}
	J= \sum_{n=1}^{n_t} \|a(t+n\delta t)-a^*\|_Q^2 + \sum_{n=0}^{n_t-1} R\, \dot{\alpha}(t+n\delta t)^2
	\label{eq:mpc_cost}
\end{equation}
subject to the constraints
\begin{align}
	\alpha_{\min} &\leq \alpha(t+n\delta t) \leq \alpha_{\max}, \quad n = 1, \dots, n_t, \label{eq:contr_alpha}\\
	\dot{\alpha}_{\min} &\leq \dot{\alpha}(t+n\delta t) \leq \dot{\alpha}_{\max}, \quad n = 0, \dots, n_t-1. \label{eq:contr_dalpha}
\end{align}
Here $Q\in\mathbb S_{++}^r$ is a weighting matrix on the deviation from the set point and $R>0$ penalises the control effort, i.e. the variations of the angle of attack.

\begin{proposition}[Convexity]\label{prop:qp}
Problem \eqref{eq:mpc_cost}--\eqref{eq:contr_dalpha}, with the dynamics \eqref{eq:augmente}, is a strictly convex quadratic program in the control sequence $U=(\dot\alpha(t),\ldots,\dot\alpha(t+(n_t-1)\delta t))^\top\in\R^{n_t}$. It therefore has at most one minimiser, which is the global minimum.
\end{proposition}

\begin{proof}
By induction, $\tilde x(t+n\delta t)=\tilde A^n\tilde x(t)+\sum_{k=0}^{n-1}\tilde A^{n-1-k}\tilde B\,\dot\alpha(t+k\delta t)$. The state predicted over the whole horizon can thus be written $X=\Phi\,\tilde x(t)+\Gamma\,U$, where $\Phi$ and $\Gamma$ are constant matrices depending only on $\tilde A$, $\tilde B$ and $n_t$. With $E=[I_r\ 0]$ extracting $a$ from $\tilde x$ and $\bar Q=I_{n_t}\otimes Q$, the criterion \eqref{eq:mpc_cost} becomes
\[
J(U)=\big(\bar E X-\bar a^*\big)^\top \bar Q\big(\bar E X-\bar a^*\big)+R\,U^\top U = U^\top H U+2f^\top U+\mathrm{const},
\]
where $\bar E=I_{n_t}\otimes E$, $\bar a^*$ repeats $a^*$ over the horizon, $H=\Gamma^\top\bar E^\top\bar Q\bar E\,\Gamma+R\,I_{n_t}$ and $f$ is affine in $\tilde x(t)$ and $a^*$. Since $\bar Q\succeq 0$ and $R>0$, we have $H\succ 0$: $J$ is strictly convex. The constraints \eqref{eq:contr_alpha} bear on $\alpha(t+n\delta t)=\alpha(t)+\delta t\sum_{k<n}\dot\alpha(t+k\delta t)$, hence are affine in $U$, as are \eqref{eq:contr_dalpha} (box constraints). The feasible set is a convex polyhedron, which proves the claim.
\end{proof}

\subsection{Solution and application of the control}
The problem is solved at each instant over a receding horizon. Only the first control $\dot\alpha^*(t)$ of the optimal sequence is applied; once the state $a(t+\delta t)$ is obtained, the problem is solved again over the horizon $[t+\delta t,\,t+(n_t+1)\delta t]$, and so on. The steps are summarised in algorithm~\ref{alg:mpc}.

\begin{algorithm}[ht]
	\begin{algorithmic}[1]
		\State \textbf{Offline:} compute $a^*$ by gradient descent on \eqref{eq:a_star}; compute $A$, $B$ by DMDc; build $\tilde A$, $\tilde B$.
		\State $t \gets t_0$
		\While{$t < t_f$}
			\State $a(t)$ and $\alpha(t)$ known; build $\tilde{x}(t) = [a(t)^\top,\, \alpha(t)]^\top$
			\State Solve the QP: find $(v(t+n\delta t))_{n=0,\ldots,n_t-1}$ that minimises
			\[\textstyle\sum_{n=1}^{n_t} \|a(t+n\delta t)-a^*\|_Q^2 + \sum_{n=0}^{n_t-1} R\, v(t+n\delta t)^2\]
			subject to $\tilde{x}(t+(n+1)\delta t)=\tilde{A}\tilde{x}(t+n\delta t)+\tilde{B}v(t+n\delta t)$, $\alpha_{\min} \leq \alpha(t+n\delta t) \leq \alpha_{\max}$ and $\dot{\alpha}_{\min} \leq v(t+n\delta t) \leq \dot{\alpha}_{\max}$
			\State Apply $v^*(t)$ (only the first control of the horizon is kept)
			\State $t \gets t + \delta t$
		\EndWhile
	\end{algorithmic}
	\caption{Angle-of-attack control by MPC with augmented state (receding horizon).}
	\label{alg:mpc}
\end{algorithm}

The linearity of the DMDc model is thus not merely a practical simplification: it is a property that guarantees the tractability of the control problem and the reliability of the resulting control laws, whereas a Galerkin model would lead to a non-convex problem (section~\ref{sec:dmdc}).

\section{Justification of the offline computation of \texorpdfstring{$a^*$}{a*}}\label{sec:precalc_a_star}

We compare the cost of the MPC algorithm for a functional $J$ defined directly by persistence diagrams, and for a functional defined through the offline computation of $a^*$. The control horizon has $n_t$ time steps, and one optimisation problem is solved at each step.

\subsection{Cost if \texorpdfstring{$J$}{J} is based on the persistence diagram}
If one directly minimises
\[
J= \sum_{n=1}^{n_t} \dW\big(\dgm(|\omega(a(t+n\delta t))|), \dgm(0)\big)^2 + \gamma \sum_{n=0}^{n_t-1} \|\boldsymbol{\alpha}(t+n\delta t)\|^2,
\]
each evaluation of $J$ requires reconstructing $\omega(a)\in\R^N$ from $a\in\R^r$, with $r\ll N$, which brings the problem back to the full physical space. Two major drawbacks follow.
\begin{enumerate}
	\item \textbf{Loss of convexity.} The composition
	\[
	\boldsymbol{\alpha}\longrightarrow a\longrightarrow\omega(a)\in\R^N\longrightarrow\dgm(|\omega(a)|)\longrightarrow\dW^2
	\]
	is non-convex in $\boldsymbol{\alpha}$, even though the dynamics \eqref{eq:dmdc} is linear. The convex quadratic structure of the MPC is lost; a nonlinear MPC becomes necessary \cite{grune2017nmpc,allgower2000nmpc}, with no guarantee of convergence to a global optimum.
	\item \textbf{Loss of the benefit of the reduced model.} If the nonlinear solver performs $K_{\mathrm{NL}}$ evaluations of $J$ per time step, the total cost over $n_t$ steps is
	\begin{equation}
		\text{Total cost} = n_t \times K_{\mathrm{NL}} \times C_{\mathrm{reconstruction}}(N),
		\label{eq:cost_no_astar}
	\end{equation}
	where $C_{\mathrm{reconstruction}}(N)$ is the cost of a reconstruction in $\R^N$. The benefit of the reduced model is thus largely lost.
\end{enumerate}

\subsection{Cost with offline computation of \texorpdfstring{$a^*$}{a*}}
The offline computation of $a^*$ consists in solving problem \eqref{eq:a_star} \emph{only once} by a gradient descent of $K'$ iterations, each involving a reconstruction in $\R^N$. Afterwards, the MPC criterion becomes
\[
J= \sum_{n=1}^{n_t} \|a(t+n\delta t)-a^*\|^2 + \gamma\sum_{n=0}^{n_t-1} \|\boldsymbol{\alpha}(t+n\delta t)\|^2,
\]
and no longer requires any reconstruction in $\R^N$. The problem is convex quadratic (proposition~\ref{prop:qp}) and is solved by a dedicated QP solver. The total cost is
\begin{equation}
	\text{Total cost} = \underbrace{K' \times C_{\mathrm{reconstruction}}(N)}_{\text{once}} + \underbrace{n_t \times K_{\mathrm{QP}} \times C_{\mathrm{QP}}(r)}_{\text{MPC loop in }\R^r},
	\label{eq:cost_with_astar}
\end{equation}
where $K_{\mathrm{QP}}$ is the number of iterations of the QP solver at a given instant and $C_{\mathrm{QP}}(r)=O(r^2)$ the cost of one iteration.

Since $r\ll N$, we have $C_{\mathrm{QP}}(r)\ll C_{\mathrm{reconstruction}}(N)$. Assuming that $K_{\mathrm{NL}}$, $K'$ and $K_{\mathrm{QP}}$ are of the same order of magnitude, the second term of \eqref{eq:cost_with_astar} is dominated by \eqref{eq:cost_no_astar}. As soon as $n_t\gg K'/K_{\mathrm{NL}}$, the first term is dominated as well. We thus obtain
\[
n_t \times K_{\mathrm{NL}} \times C_{\mathrm{reconstruction}}(N) \gg K' \times C_{\mathrm{reconstruction}}(N) + n_t \times K_{\mathrm{QP}} \times C_{\mathrm{QP}}(r).
\]
The proposed approach therefore offers a fast and convex control strategy, compatible with real-time application.

\section{Numerical results}\label{sec:resultats}

\subsection{Set-up}
We apply the approach to the unsteady flow around an aircraft wing. We assume that, whatever the imposed angle of attack, the wing vibrates with a constant frequency $f$; seven frequencies are studied: $f\in\{0.10,\,0.20,\,0.25,\,0.30,\,0.35,\,0.40,\,0.50\}$\,Hz. To build $\dgm'(|\omega(a)|)$, which enters the definition of $a^*$, we remove the two most persistent points (section~\ref{sec:cible}): the control aims at reducing the vortices other than the main vortex and the secondary vortex. The reduced model is a POD--DMDc model, and the control is computed by algorithm~\ref{alg:mpc}.

\begin{figure}\small
	\centering
	\begin{tabular}{cc}
		\includegraphics[width=.4\textwidth]{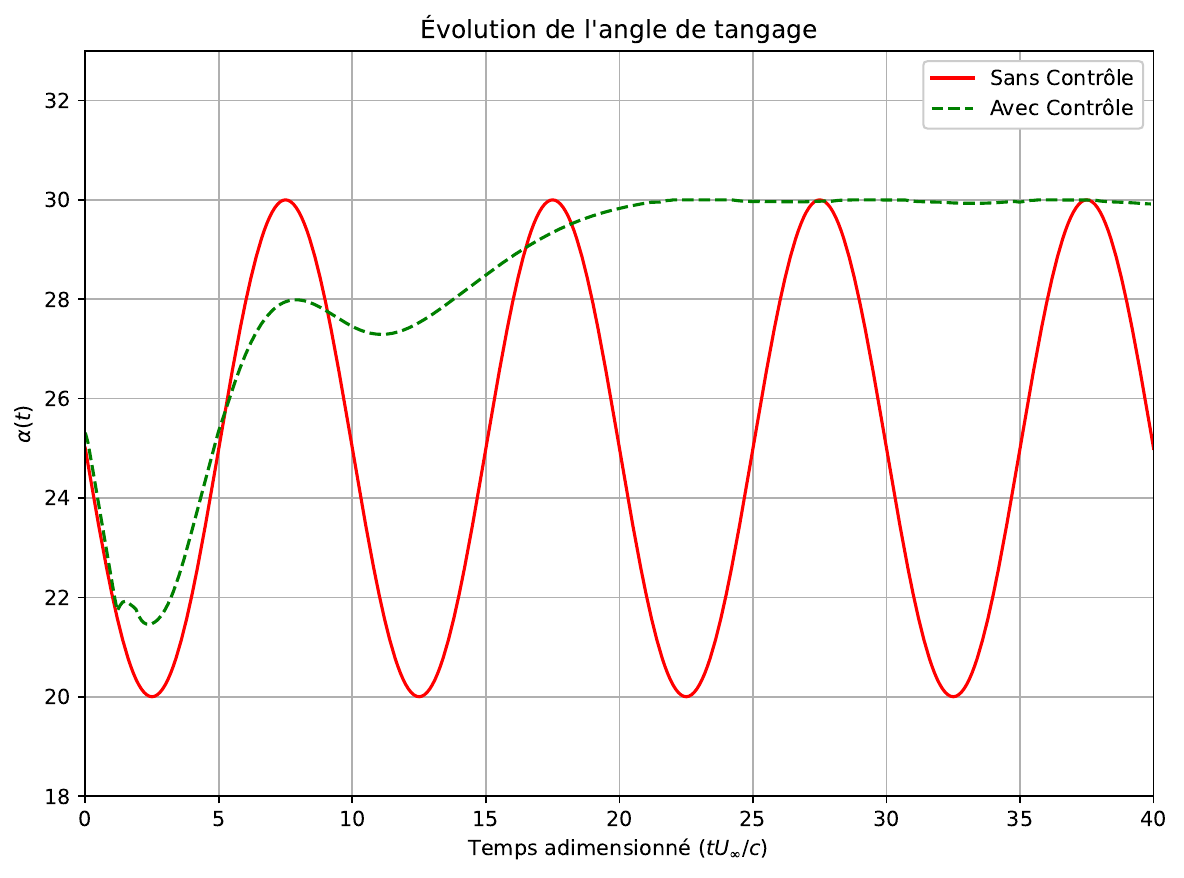}&
		\includegraphics[width=.4\textwidth]{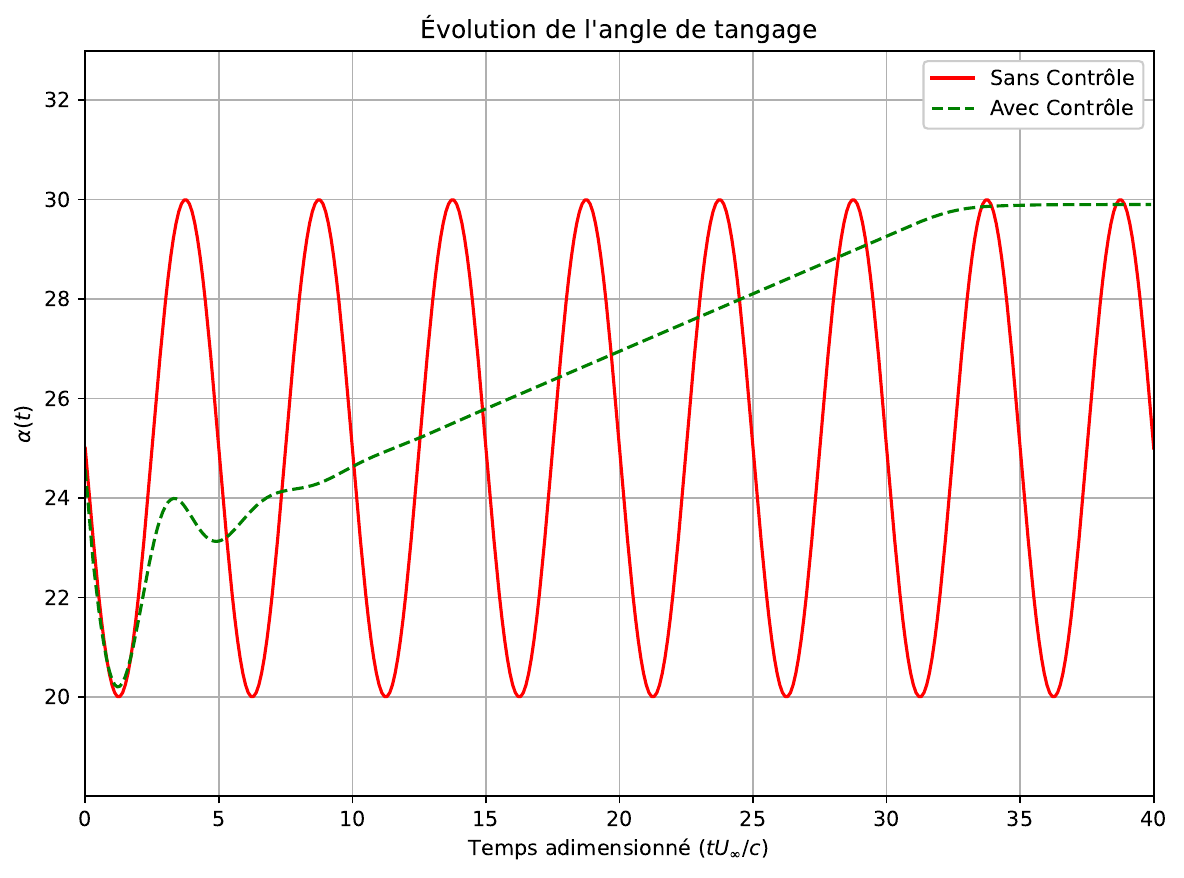}\\
		\textbf{a. }$f=0.10$&\textbf{b. }$f=0.20$\\
		\includegraphics[width=.4\textwidth]{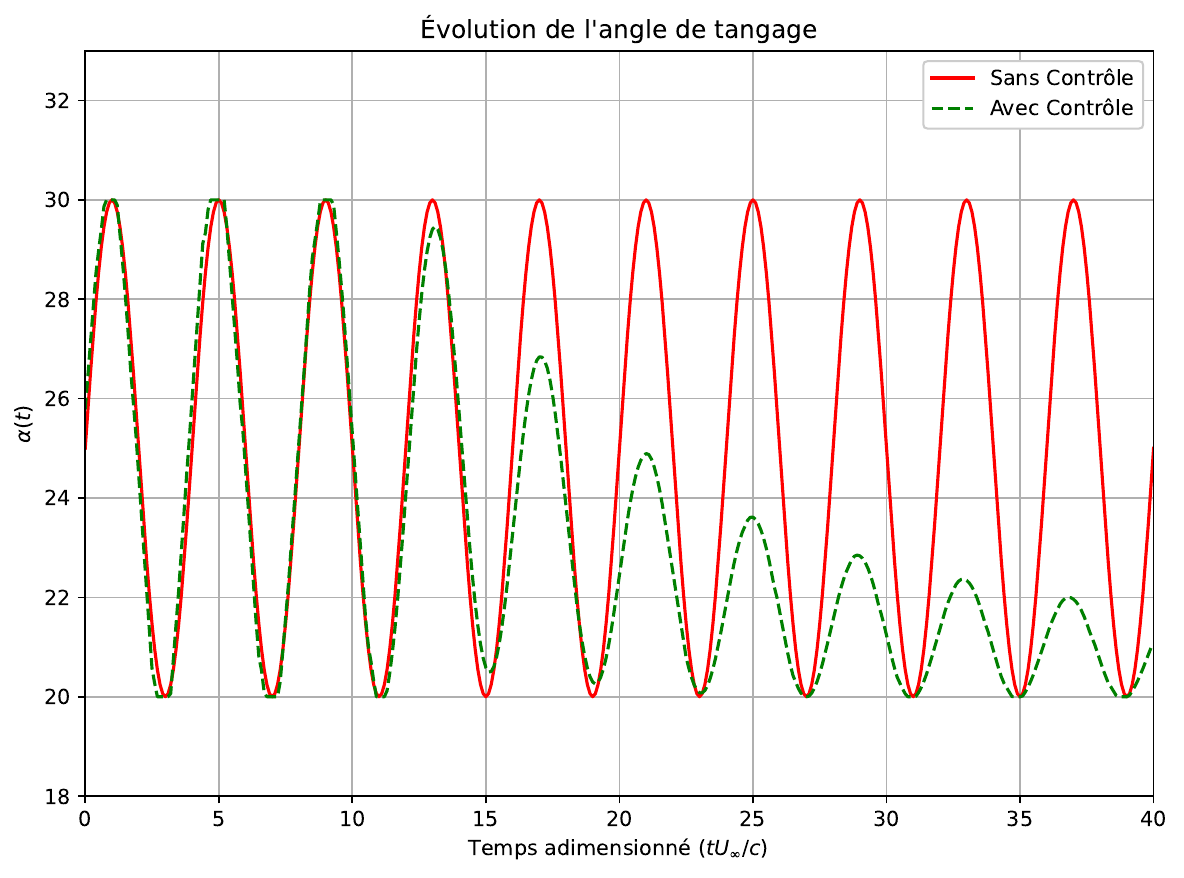}&
		\includegraphics[width=.4\textwidth]{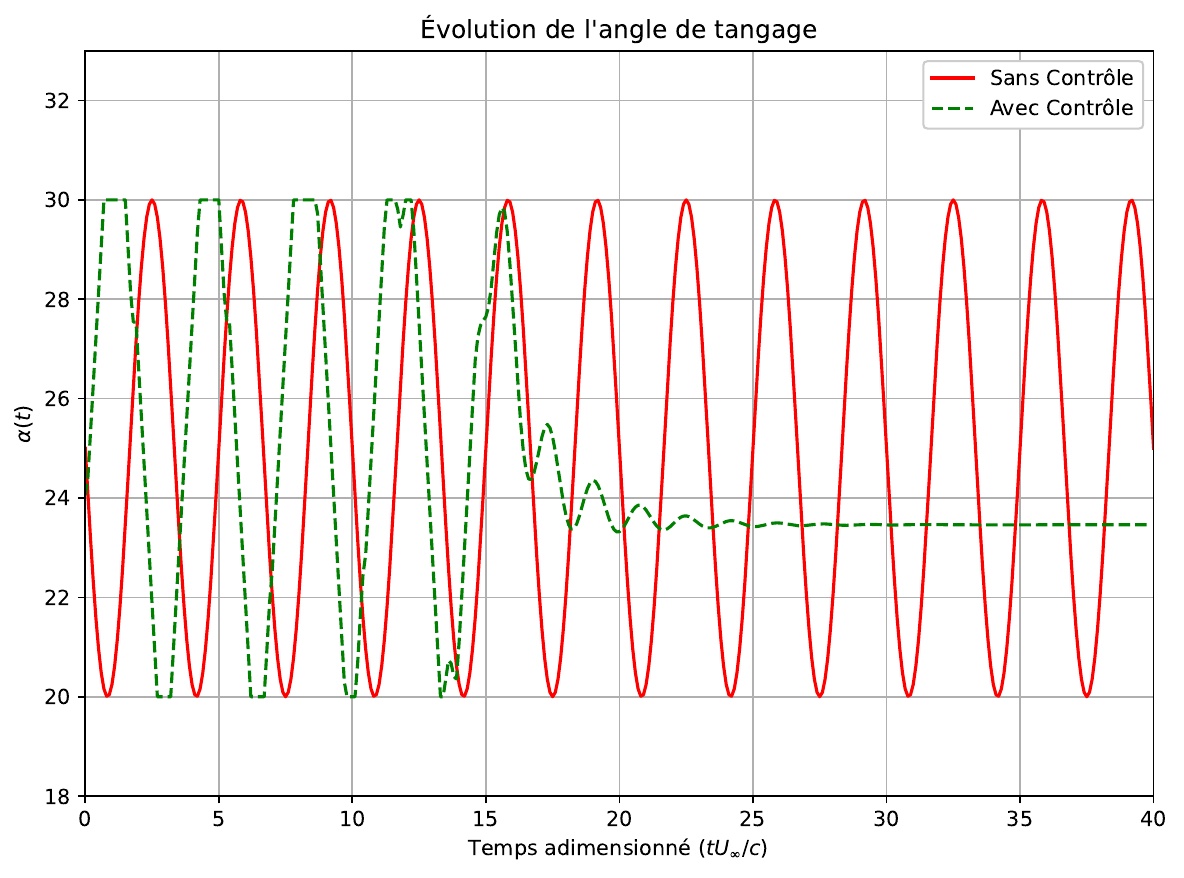}\\
		\textbf{c. }$f=0.25$&\textbf{d. }$f=0.30$\\
		\includegraphics[width=.4\textwidth]{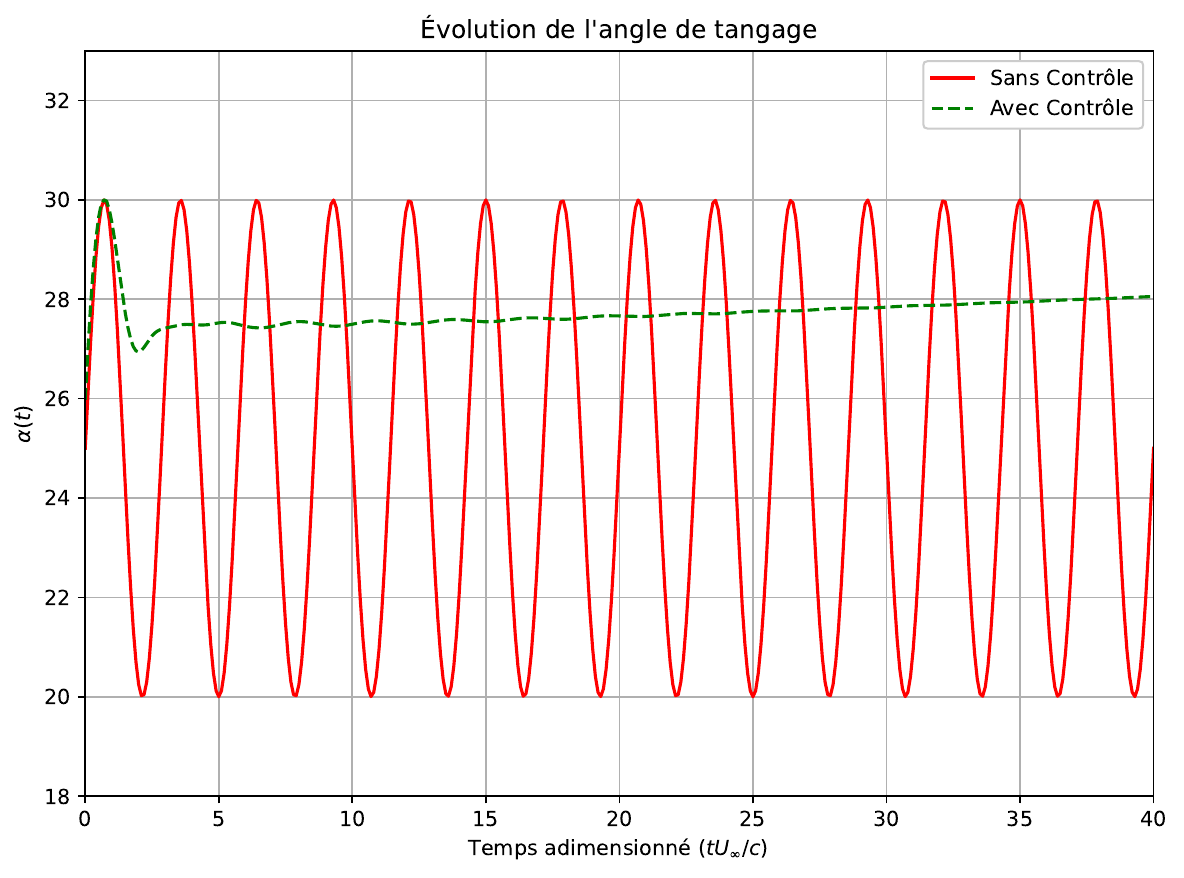}&
		\includegraphics[width=.4\textwidth]{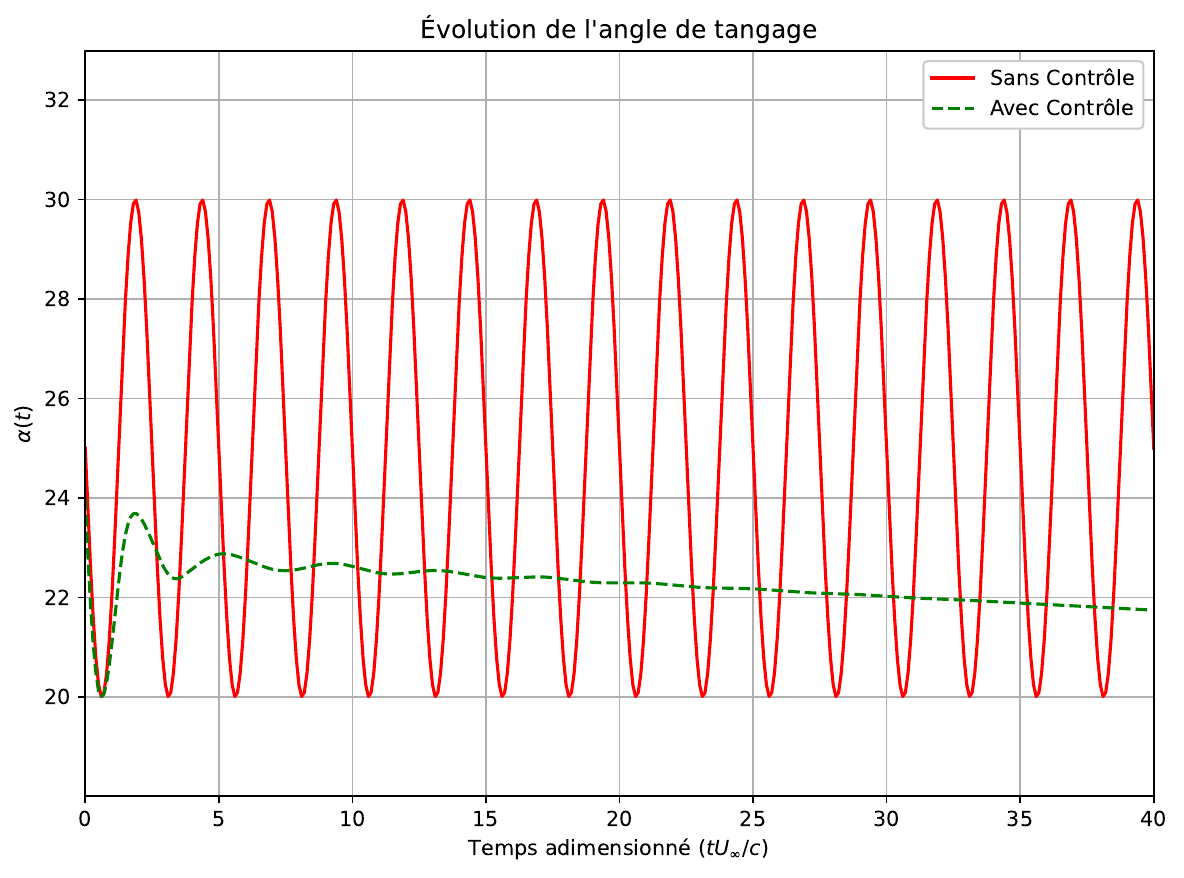}\\
		\textbf{e. }$f=0.35$&\textbf{f. }$f=0.40$\\
		\includegraphics[width=.4\textwidth]{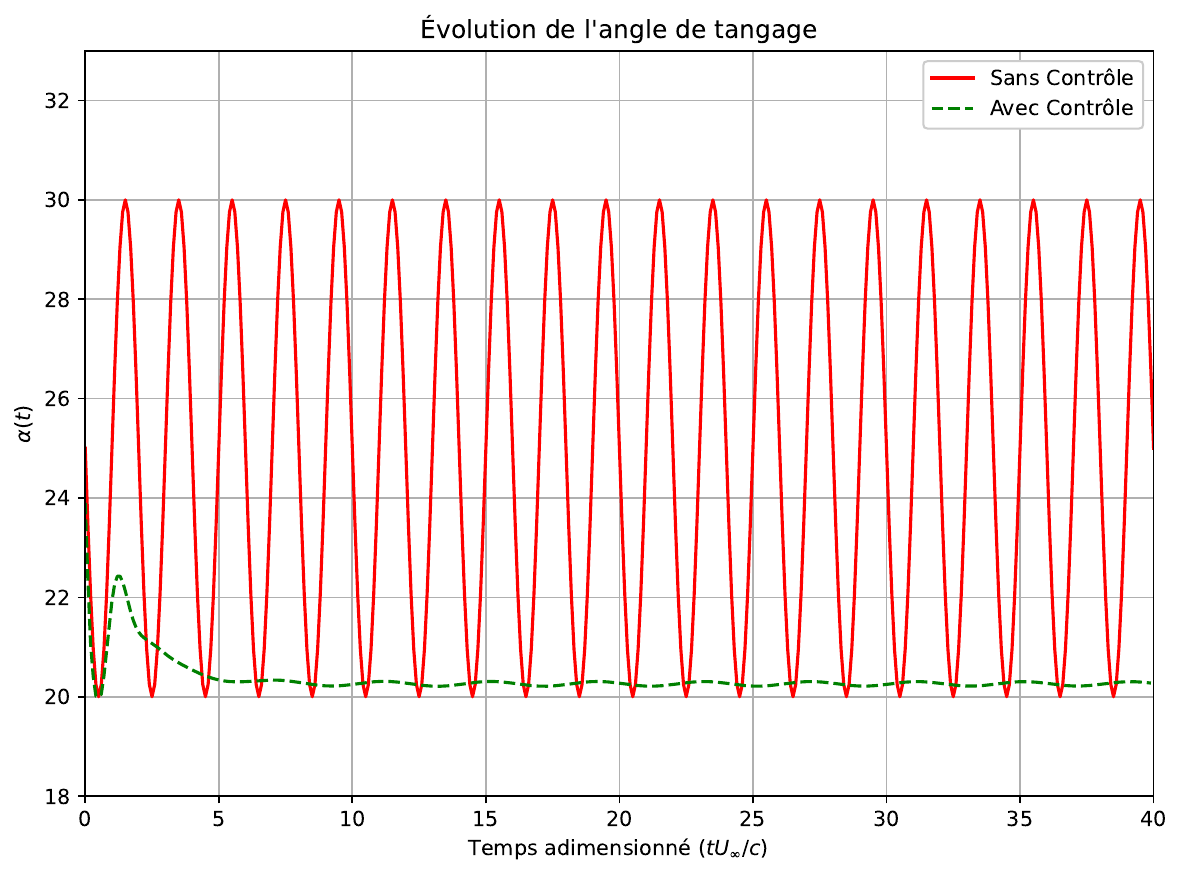}\\
		\textbf{g. }$f=0.50$\\
	\end{tabular}
	\caption{Time evolution of the pitch angle $\alpha(t)$, without and with control.}
	\label{fig:control_angle}
\end{figure}

The control-effort regularisation parameter is set to $R=0.1$. A small value of $R$ favours fast convergence to the target flow, at the cost of larger angle-of-attack variations that are closer to the bounds $\dot\alpha_{\min},\dot\alpha_{\max}$; a large value smooths the control at the expense of the speed at which undesirable vortices are reduced. The value $R=0.1$ was chosen to obtain a fast and significant drag reduction (table~\ref{tab:control_cd}) while keeping angle-of-attack variations realistic and compatible with the actuator bounds (figure~\ref{fig:control_angle}).

\subsection{Effect on the vorticity field}
\begin{figure}\small
	\centering
	\includegraphics[width=\textwidth]{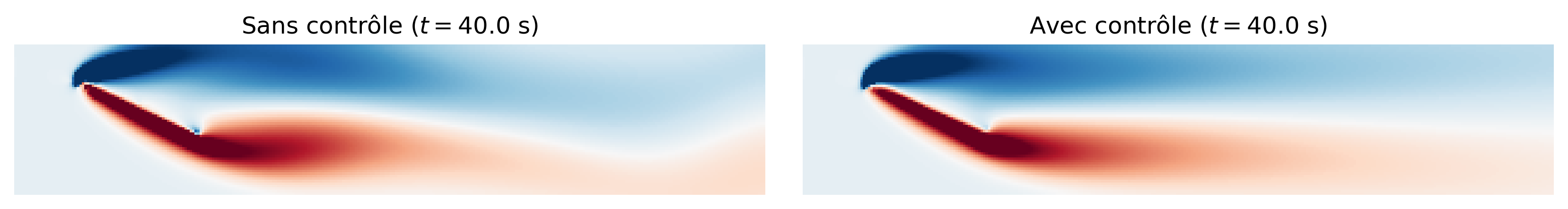}\\
	\textbf{a. }$f=0.10$\\[5pt]
	\includegraphics[width=\textwidth]{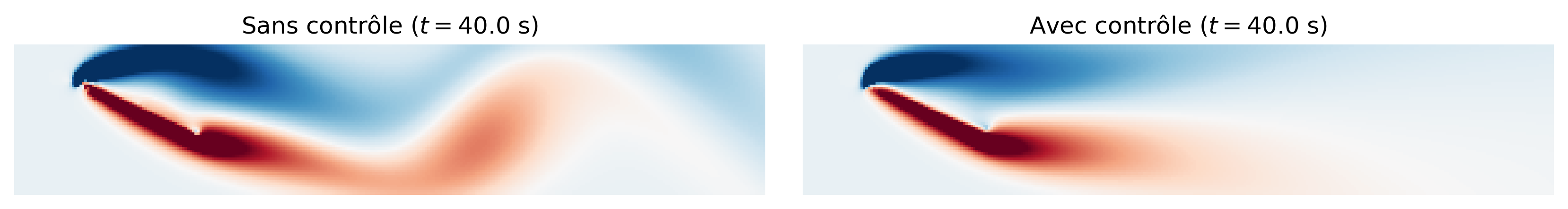}\\
	\textbf{b. }$f=0.20$\\[5pt]
	\includegraphics[width=\textwidth]{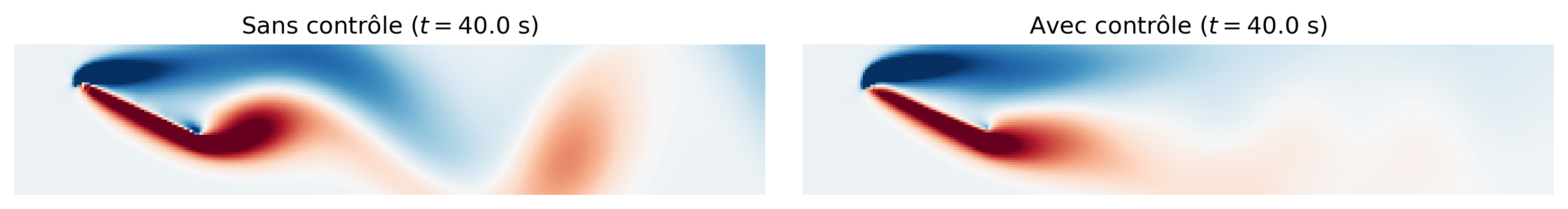}\\
	\textbf{c. }$f=0.25$\\[5pt]
	\includegraphics[width=\textwidth]{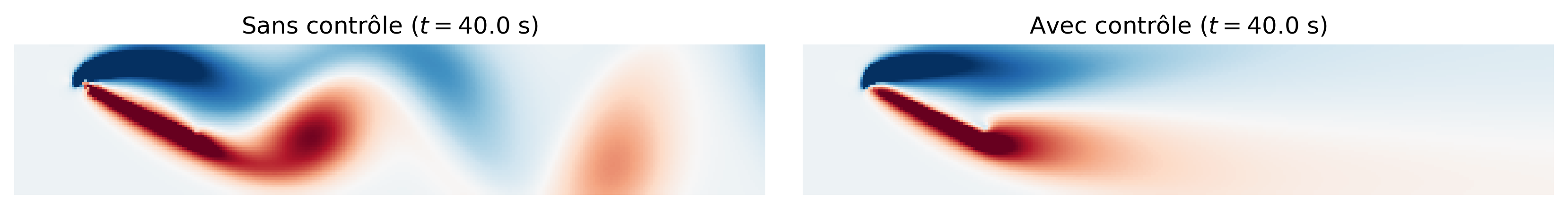}\\
	\textbf{d. }$f=0.30$\\[5pt]
	\includegraphics[width=\textwidth]{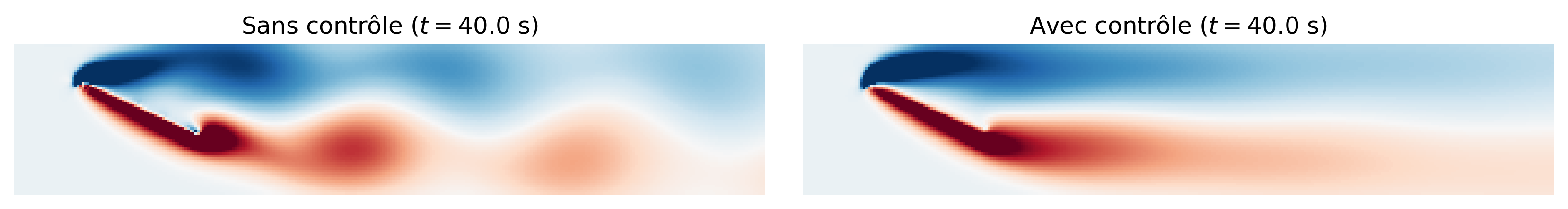}\\
	\textbf{e. }$f=0.35$\\[5pt]
	\includegraphics[width=\textwidth]{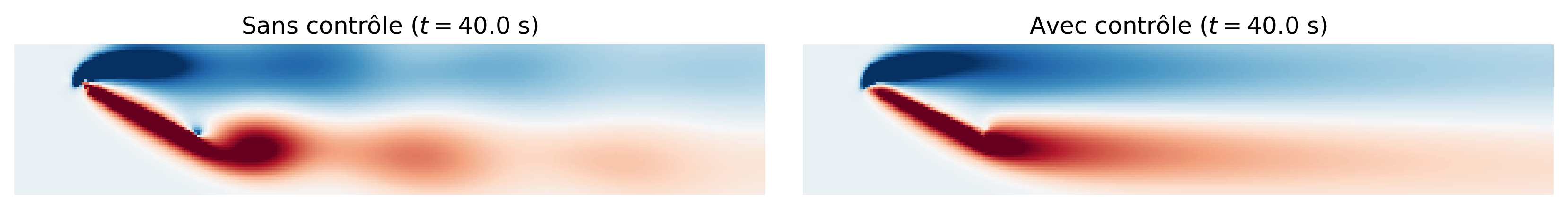}\\
	\textbf{f. }$f=0.40$\\[5pt]
	\includegraphics[trim={0cm 65mm 0cm 0cm},clip,width=\textwidth]{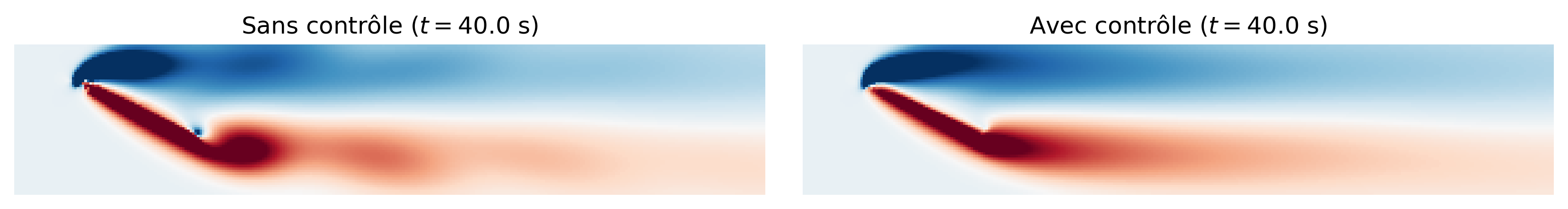}\\
	\textbf{g. }$f=0.50$\\[5pt]
	\includegraphics[trim={0cm 5mm 0cm 75mm},clip,width=\textwidth]{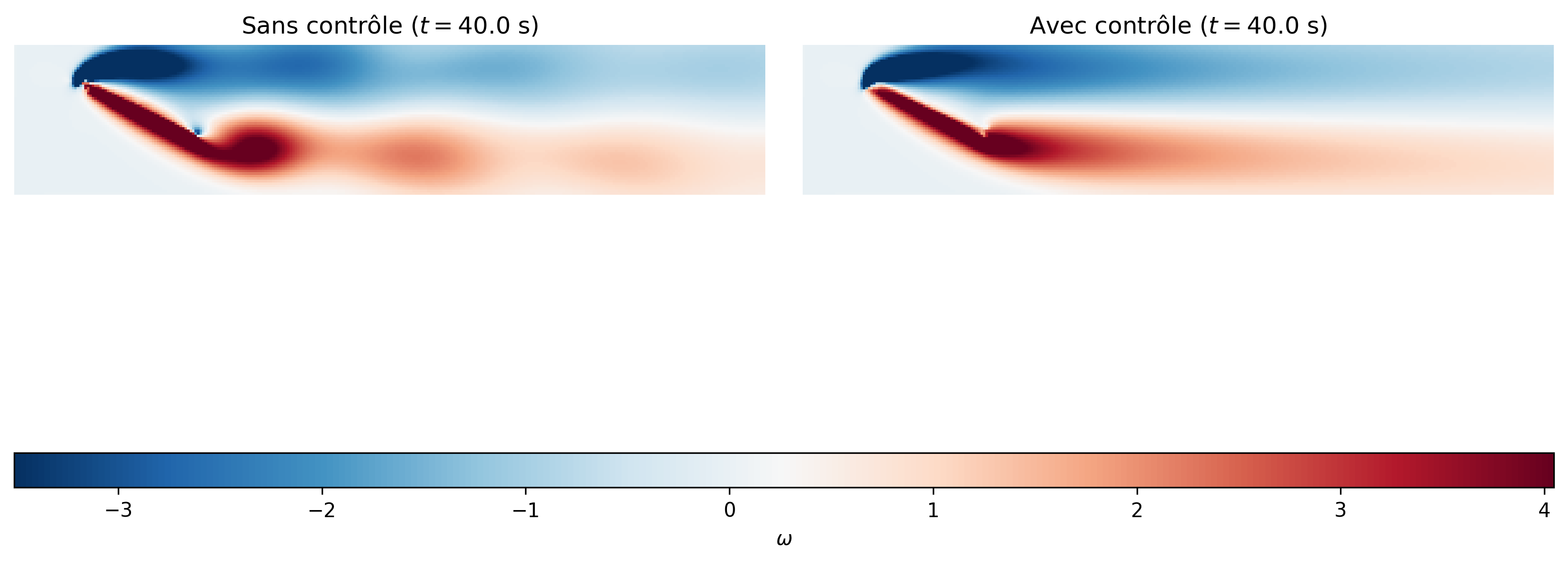}\\
	\textbf{h. }Legend ($\omega$)
	\caption{Effect of the control on the vorticity field at $t=40$\,s.}
	\label{fig:control_vorticity}
\end{figure}

Figure~\ref{fig:control_vorticity} shows the vorticity field at the end of the simulation ($t=40$\,s), with and without control. The control keeps the two most intense vortices, even though their shapes change: in the vast majority of cases they shrink, but the maximum vorticity varies little. In contrast, the control sharply reduces the intensity of the other vortices. For $f=0.10$\,Hz, there are no vortices other than the two main ones in the region of interest; the control then reduced the vorticity intensity in the region downstream of the wing. The time evolution of the vorticity field can be viewed at {\small\url{https://yiphta-lab.github.io/videos-these/}}.

\subsection{Drag reduction}
\begin{figure}\small
	\centering
	\begin{tabular}{cc}
		\includegraphics[width=.46\textwidth]{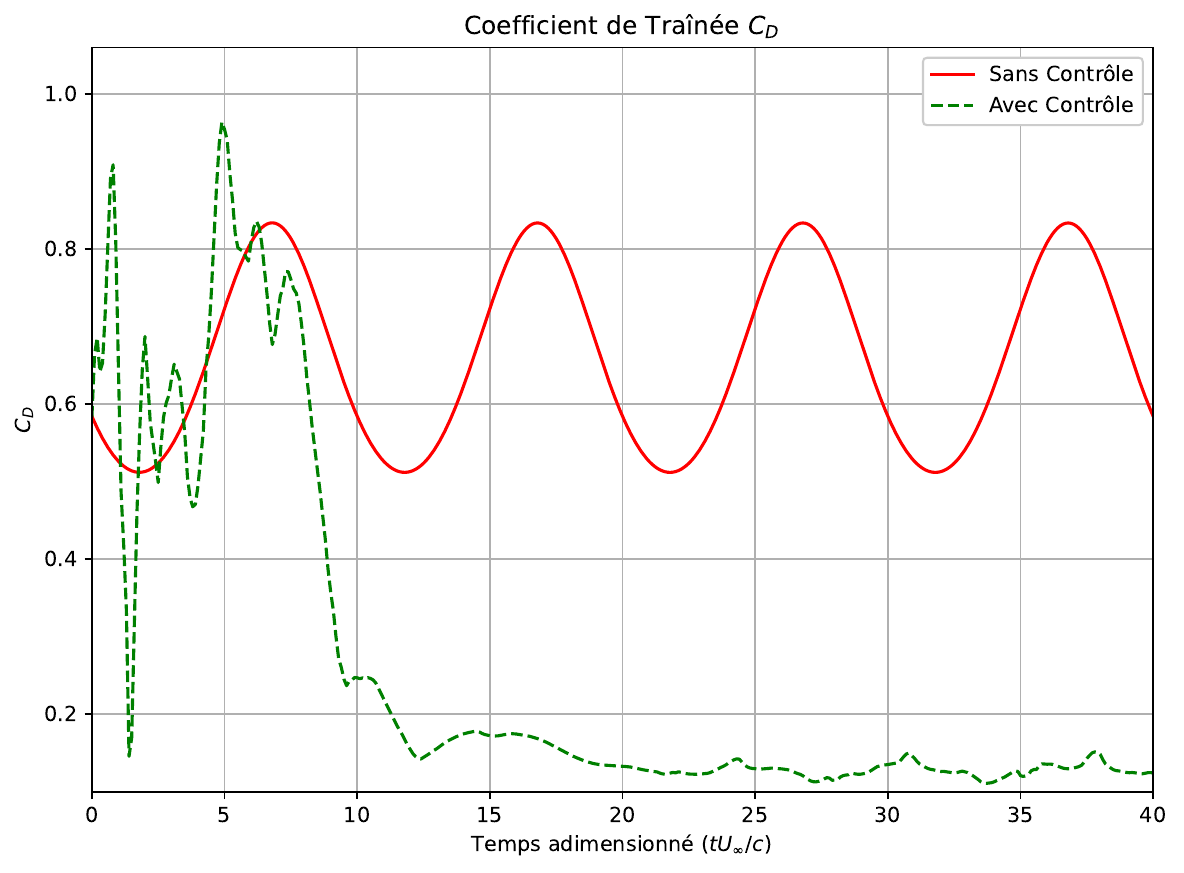}&
		\includegraphics[width=.46\textwidth]{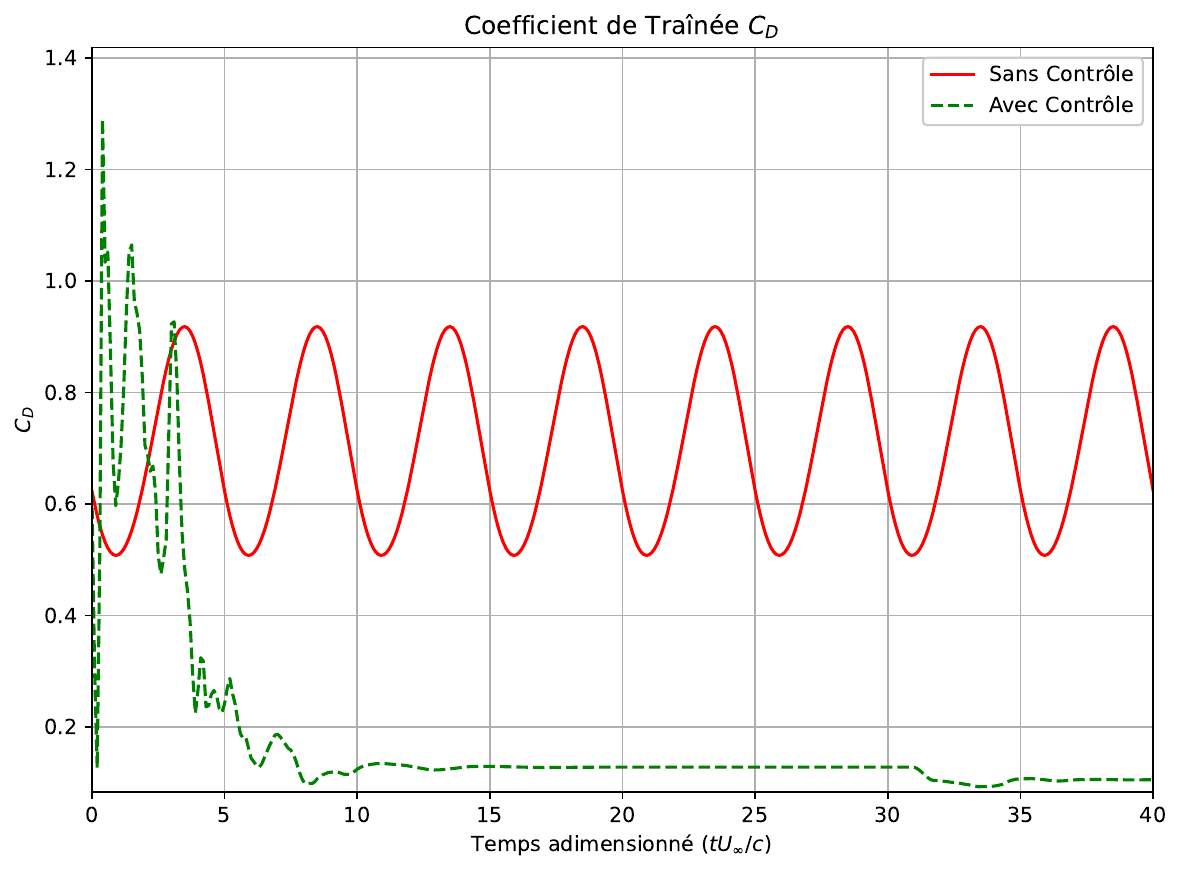}\\
		\textbf{a. }$f=0.10$&\textbf{b. }$f=0.20$\\
		\includegraphics[width=.46\textwidth]{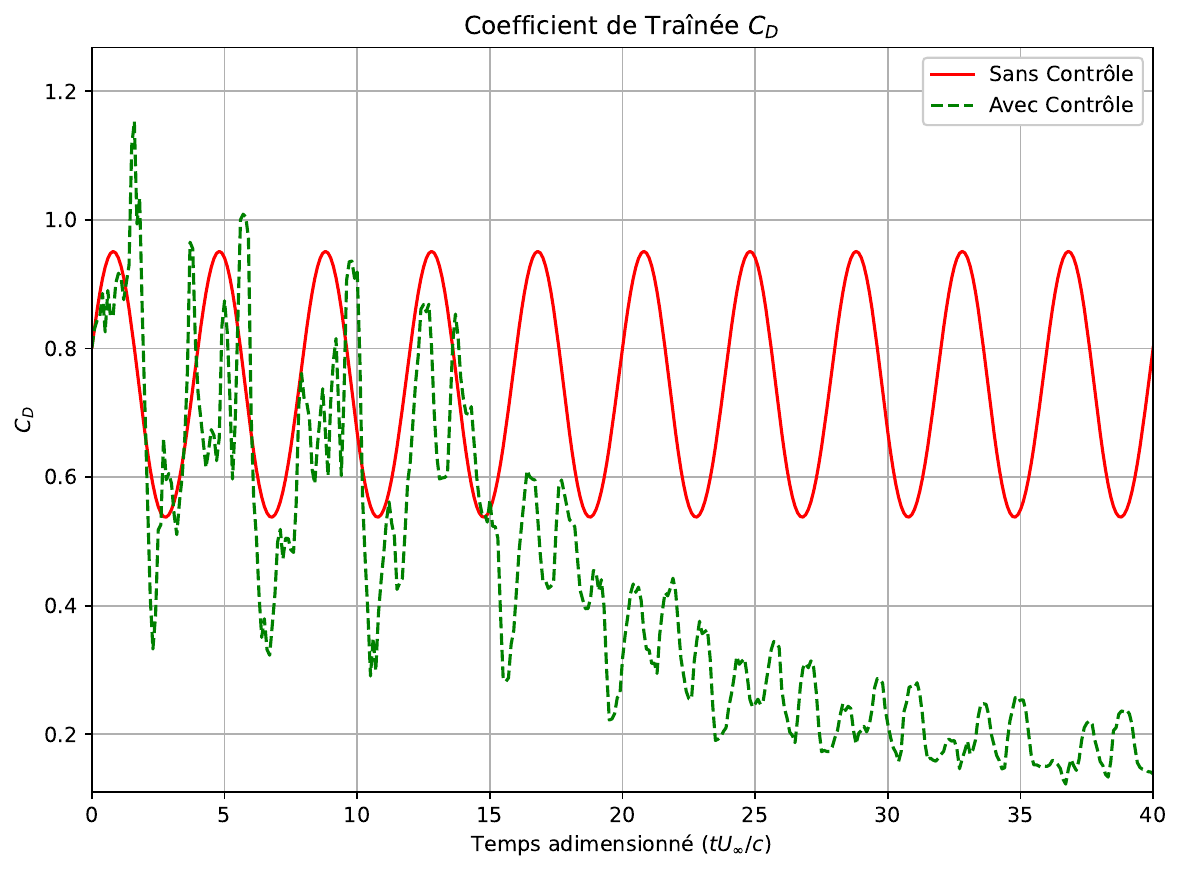}&
		\includegraphics[width=.46\textwidth]{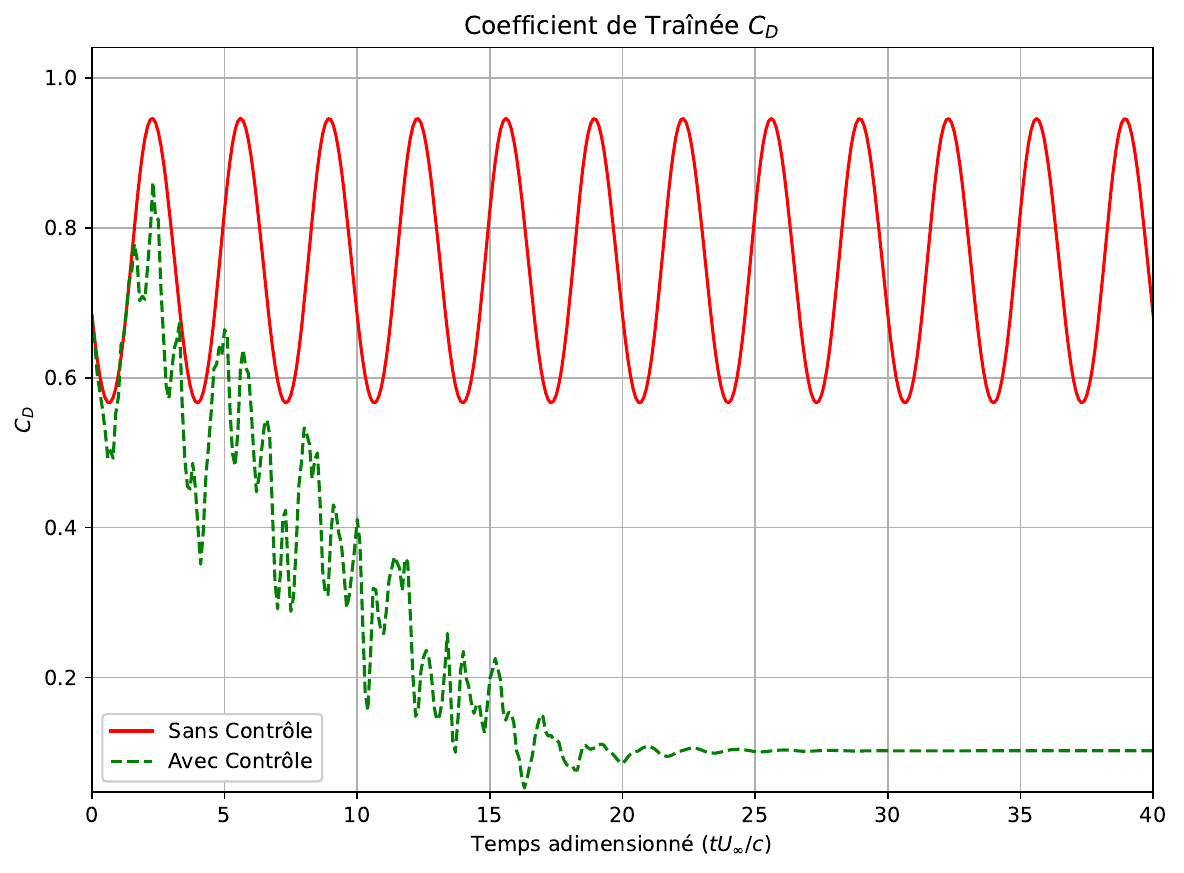}\\
		\textbf{c. }$f=0.25$&\textbf{d. }$f=0.30$\\
		\includegraphics[width=.46\textwidth]{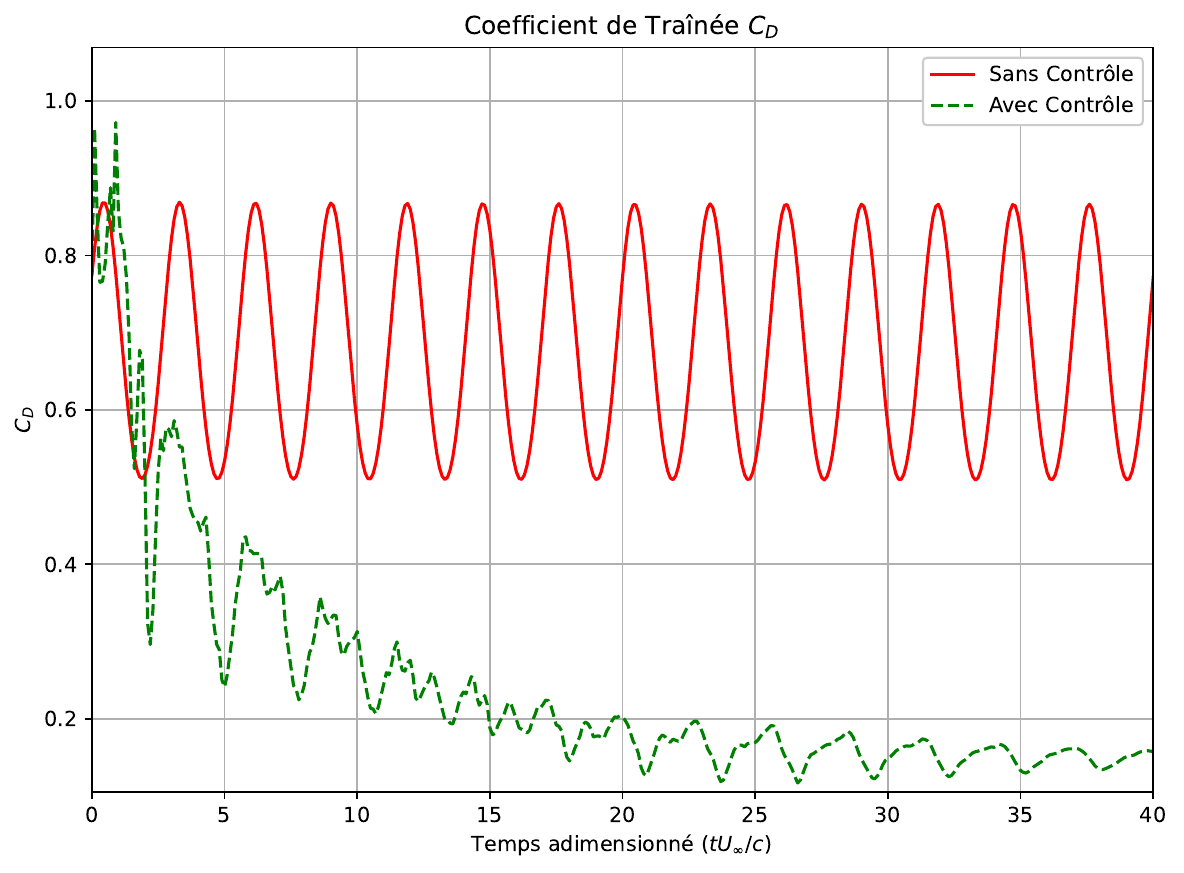}&
		\includegraphics[width=.46\textwidth]{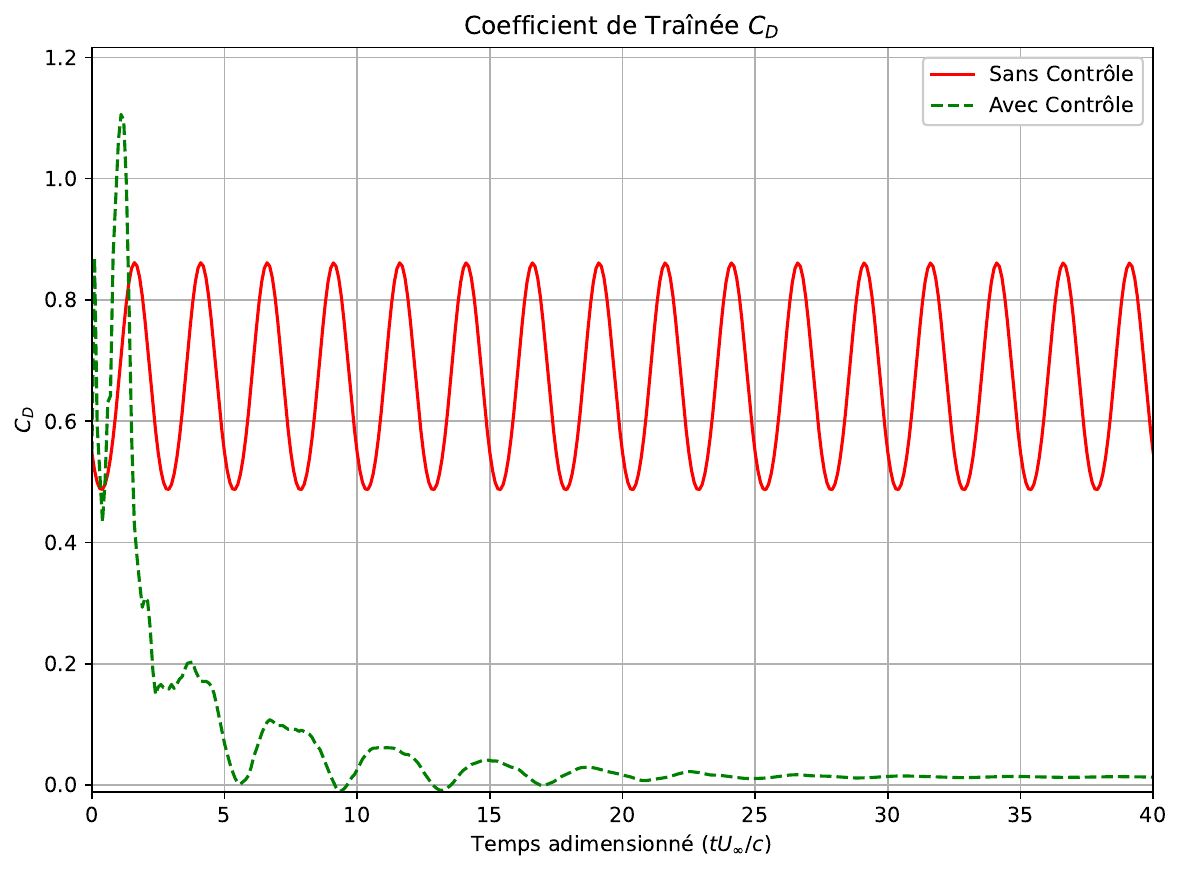}\\
		\textbf{e. }$f=0.35$&\textbf{f. }$f=0.40$\\
		\includegraphics[width=.46\textwidth]{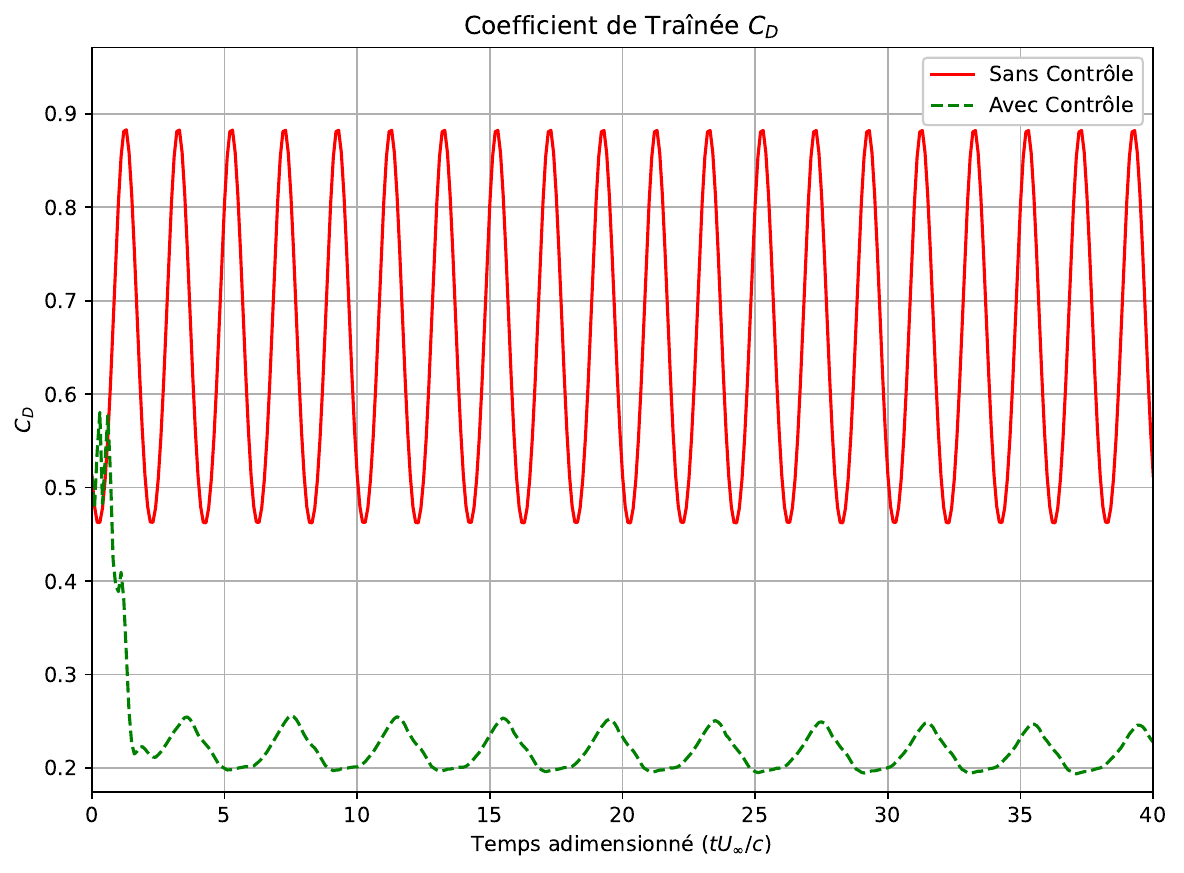}\\
		\textbf{g. }$f=0.50$\\
	\end{tabular}
	\caption{Effect of the control on the drag coefficient $C_d$.}
	\label{fig:control_cd1}
\end{figure}

Figure~\ref{fig:control_cd1} shows the evolution of the drag coefficient $C_d$. Without control, $C_d$ is periodic and oscillates at the same frequency $f$ as the wing vibration. With control, it decreases drastically within a rather short time, and the oscillations are strongly attenuated for most frequencies.

Table~\ref{tab:control_cd} gives the mean of $C_d$ over the second half of the simulation, $t\in[20,40]$\,s. The decrease is significant for all frequencies. On average, $C_d$ goes from $0.7056$ without control to $0.1289$ with control, a reduction of about $82\,\%$ ($1-0.1289/0.7056\approx 0.817$).

\begin{table}[ht]
	\centering
	\begin{tabular}{l|ccccccc|c}
		\toprule
		Frequency $f$ (Hz) & 0.10 & 0.20 & 0.25 & 0.30 & 0.35 & 0.40 & 0.50 & Mean \\
		\midrule
		Without control & 0.674 & 0.822 & 0.619 & 0.847 & 0.611 & 0.706 & 0.660 & \textbf{0.7056} \\
		With control & 0.126 & 0.105 & 0.179 & 0.102 & 0.149 & 0.014 & 0.227 & \textbf{0.1289} \\
		\bottomrule
	\end{tabular}
	\caption{Mean drag coefficient $C_d$ for $t\in[20,40]$\,s.}
	\label{tab:control_cd}
\end{table}

\section{Discussion, limitations and perspectives}\label{sec:discussion}

The control strategy is embedded in a POD--DMDc reduced model; it therefore inherits its advantages (speed, linearity) but also its limitations. In particular, the model is built from snapshots obtained for a given range of variation of the angle of attack. If the angle leaves this range, the model is no longer valid and the control may fail; the bounds $\alpha_{\min},\alpha_{\max}$ of the constraints \eqref{eq:contr_alpha} must therefore be chosen within its domain of validity. An adaptive reduced model, updated with data acquired during control, is a natural direction.

The global optimality established by proposition~\ref{prop:qp} concerns the MPC problem posed in the reduced space, with a linear model, a quadratic cost and affine constraints. The computation of $a^*$ by gradient descent on \eqref{eq:a_star} is, on the other hand, a non-convex problem: it yields a local minimum, whose quality depends on the initialisation. The non-convexity is however confined to an offline step, executed only once.

The definition of $\dgm'$ can be adapted to the application (removal of $k$ points, persistence threshold). However, the diagram alone cannot select a vortex according to its position in space; combining the diagram with the mergegram \cite{elkin2020mergegram} or other spatial descriptors is an interesting perspective. Likewise, replacing diagrams by persistence images or other vector representations could simplify the computation of gradients.

Finally, the results concern seven vibration frequencies of a single wing configuration, with a single value of $R$. Sensitivity studies (weights $Q$ and $R$, horizon length $n_t$, number of modes $r$, number of points removed from the diagram) and tests on other geometries or flow regimes remain to be carried out. The effect of the control on lift should be quantified, the preservation of leading-edge structures being observed here only through the vorticity fields.

\section{Conclusion}\label{sec:conclusion}

We have presented a flow-control approach aimed at selectively reducing undesirable vortices through topological data analysis. The target flow is characterised by a modified persistence diagram, and its offline computation, by gradient descent with a diagram derivative defined through a parametrisation, brings the control problem back to the reduced space. MPC with augmented state on a linear DMDc model leads to a strictly convex quadratic program, whose solution is a global optimum, and whose cost is much lower than that of an approach using diagrams inside the control loop. Applied to the flow around an oscillating wing, the method reduces the drag coefficient by about $82\,\%$ on average while keeping the main vortices. These results open the way to real-time applications for the control of complex flows, provided the control remains within the domain of validity of the reduced model.

\bibliographystyle{plain}
\bibliography{mbangue}

\end{document}